\documentclass{amsart}
\usepackage{mathrsfs}
\usepackage{amssymb}
\usepackage{graphicx}

\def\eqlaw{\stackrel{\mbox{\tiny (law)}}{=}}     
\newcommand{\deriv}[2]{\frac{\mathrm{d}#1}{\mathrm{d}#2}}
\newcommand{\derivp}[2]{\frac{\partial #1}{\partial #2}}

\usepackage[usenames,dvipsnames]{color}

\theoremstyle{definition}

\newtheorem{example}{Example}

\theoremstyle{remark}
\newtheorem{remark}{Remark}[section]

\numberwithin{equation}{section}

\def\Xint#1{\mathchoice
{\XXint\displaystyle\textstyle{#1}}%
{\XXint\textstyle\scriptstyle{#1}}%
{\XXint\scriptstyle\scriptscriptstyle{#1}}%
{\XXint\scriptscriptstyle\scriptscriptstyle{#1}}%
\!\int}
\def\XXint#1#2#3{{\setbox0=\hbox{$#1{#2#3}{\int}$}
\vcenter{\hbox{$#2#3$}}\kern-.5\wd0}}

\def\dashint{\Xint-}

\def\d{\text{d}}
\def\e{\text{e}}
\def\i{\text{i}}

\begin{document}

\title[Supersymmetry and disorder]
  {Supersymmetric quantum mechanics with L\'{e}vy disorder in one dimension} 

\author{Alain Comtet}
\address{UPMC Univ. Paris 6, 75005 Paris and
  Univ. Paris Sud ; CNRS ; LPTMS, UMR 8626, 
             Orsay F-91405, France}
\email{alain.comtet@u-psud.fr}        

\author{Christophe Texier}
\address{Univ. Paris Sud ; CNRS ; LPTMS, UMR 8626 \& LPS, UMR 8502,
             Orsay F-91405, France}
  \email{christophe.texier@u-psud.fr}        

\author{Yves Tourigny}
\address{School of Mathematics\\
        University of Bristol\\
        Bristol BS8 1TW, United Kingdom}
  \email{y.tourigny@bristol.ac.uk}

\subjclass{Primary 82B44. Secondary 60G51}



\begin{abstract}
We consider the Schr\"{o}dinger equation with a random
potential of the form
$$
V(x) = \frac{w^2(x)}{4}-\frac{w'(x)}{2}
$$
where $w$ is a L\'{e}vy noise. We focus on the problem of computing the so-called complex Lyapunov exponent
$$
\Omega := \gamma - \i \pi N 
$$
where $N$ is the integrated density of states of the system, and $\gamma$ is the Lyapunov exponent.
In the case where the L\'{e}vy process is non-decreasing, we show that
the calculation of $\Omega$ reduces to a Stieltjes moment problem, we ascertain the low-energy behaviour
of the density of states in some generality, and relate it to the distributional properties of the L\'evy process.
We review the known solvable cases---where 
$\Omega$ can be expressed in terms of special functions--- and discover a new one. 
\end{abstract} 

\thanks{This work was supported by ``Triangle de la Physique''.}

\maketitle

\section{Introduction}
\label{introductionSection}
The disordered model we consider
is a particular case of the Schr\"{o}dinger equation
\begin{equation}
  - \psi'' + V(x) \,\psi = E \psi\,.
  \label{schroedingerEquation}
\end{equation}
In this expression, $\psi$ is the unknown (wave) function of the independent variable $x$, $E$ is the energy parameter, and $V$
is the random potential.
If $V$ has stationary increments, then
the most accessible quantities describing the model's behaviour are the {\em integrated density of states} $N(E)$, which counts
the number of energy levels below $E$ per unit length, and the 
{\em Lyapunov exponent} $\gamma(E)$, whose reciprocal provides a measure of the localisation length at level $E$.
These are conveniently expressed in terms of the limit 
\begin{equation*}
  \lim_{x \rightarrow \infty} \frac{\ln \psi (x,E)}{x}
\end{equation*}
where $\psi(\cdot,E)$ is the particular solution of Equation (\ref{schroedingerEquation}) satisfying $\psi(0,E) = 0$ and $\psi'(0,E)=1$.
This limit is a self-averaging (non-random) quantity whose almost-sure value we denote by
$\Omega(E)$. The relationship between the real numbers $\gamma(E)$ and $N(E)$ and the complex number $\Omega(E)$ is then (see \cite{Lu,Nie82,Ni})
\begin{equation}
\Omega(E) = \gamma(E)  - \i \pi N(E)\,.
\label{characteristicFunction}
\end{equation}
We call $\Omega$ the {\em complex Lyapunov exponent},
or the characteristic function,  
of the disordered system (\ref{schroedingerEquation}). 

Building on the pioneering work of Frisch \& Lloyd \cite{FL}, Kotani \cite{Ko} made a rigorous and detailed study of
the integrated density of states when the potential is a {\em L\'{e}vy noise}. In the present paper, we shall instead
examine the case where
\begin{equation}
V(x) = \frac{w^2(x)}{4}-\frac{w'(x)}{2}
\label{supersymmetricPotential}
\end{equation}
and it is the {\em superpotential} $w$--- rather than $V$ itself--- that is the L\'{e}vy noise. 
Historically, the study of the random supersymmetric case was initiated by Ovchinnikov \& Erikhman \cite{OE} in the context of one-dimensional disordered 
semiconductors. 
The same model, expressed later on in terms of Equations (\ref{dirac1}) and (\ref{dirac2}), may also be derived as the square of a Dirac operator with a 
random mass. As such, it is relevant in several contexts of condensed matter 
physics, including one-dimensional disordered metals, random spin chains and
organic conductors; see \cite{CT,TexHag10} for recent reviews. 
From the point of view of Anderson localisation, the one-dimensional
case is special, 
for localisation takes place as soon as there is {\em any} disorder
\cite{GMP,KS},
in contrast with the richer higher-dimensional case 
where a localisation transition can occur.  
Nevertheless, there are compelling reasons
for undertaking this study: in the deterministic case, supersymmetry permits a detailed mathematical treatment of the spectral problem, leading in a
few cases to exact results \cite{CKS}. It is therefore natural to ask
whether this favourable state of affairs persists in the presence of
randomness. 
Recently, we showed how to extend the Frisch--Lloyd approach to a
general class of random, point-like,
scatterers, including one of supersymmetric type \cite{CTT}, and 
one of our aims is to develop that preliminary work so as to bring out
the very elegant structure peculiar 
to the supersymmetric case. Another powerful motivation is the equivalence between the supersymmetric model and the mathematical description of diffusion in a random environment \cite{BCGL,Ca,Le,LeDMonFis99,Sh,TexHag09}.
In the remainder of this introduction, we review the terminology and give an outline
of our main results.

\subsection{L\'{e}vy processes}
\label{levySubsection}

Any real-valued process with right-continuous, left-limited paths started at the origin, and stationary independent increments, is called 
a {\em L\'{e}vy process} \cite{Ap1,Be}. 
Such a process, say $W$,
is completely determined
by its {\em L\'{e}vy exponent} $\Lambda$, defined implicitly by
\begin{equation}
  {\mathbb E} \left ( \e^{ \i \theta W(x) }\right ) = \e^{x \Lambda(\theta)} \,. 
\label{characteristicExponent}
\end{equation}
Furthermore, the L\'{e}vy--Khintchine formula holds:
\begin{equation}
  \Lambda(\theta) = 
  2 g\,\left [ \i \mu \, \theta - \theta^2 \right ] 
  + \int_{{\mathbb R}} 
   \left[ \e^{\i \theta y} -1 -  \frac{\i \theta y}{1+y^2} \right ] \, 
   m( \d y)
\label{levyKhintchineFormula}
\end{equation}
for some constants $\mu g \in {\mathbb R}$, $g \ge 0$, and some {\em L\'{e}vy measure} ${m}$, i.e. a measure on ${\mathbb R}$ such that
\begin{equation}
  \int_{{\mathbb R}} \min \{ 1,y^2 \} \,{m}( \d y) < \infty\,.
  \label{jumpCondition}
\end{equation}
Conversely, given numbers $\mu g$ and $g$, and a L\'{e}vy measure ${m}$,
there is a L\'{e}vy process $W$ with L\'{e}vy exponent (\ref{characteristicExponent}); 
$\mu g$, $g$ and ${m}$ are called the {\em L\'{e}vy characteristics} 
of the process.

\begin{figure}[htbp]
\vspace{7.5cm} 
\begin{picture}(0,0) 
\put(-95,0){(a)}
\put(100,0){(b)} 
\end{picture} 
\includegraphics{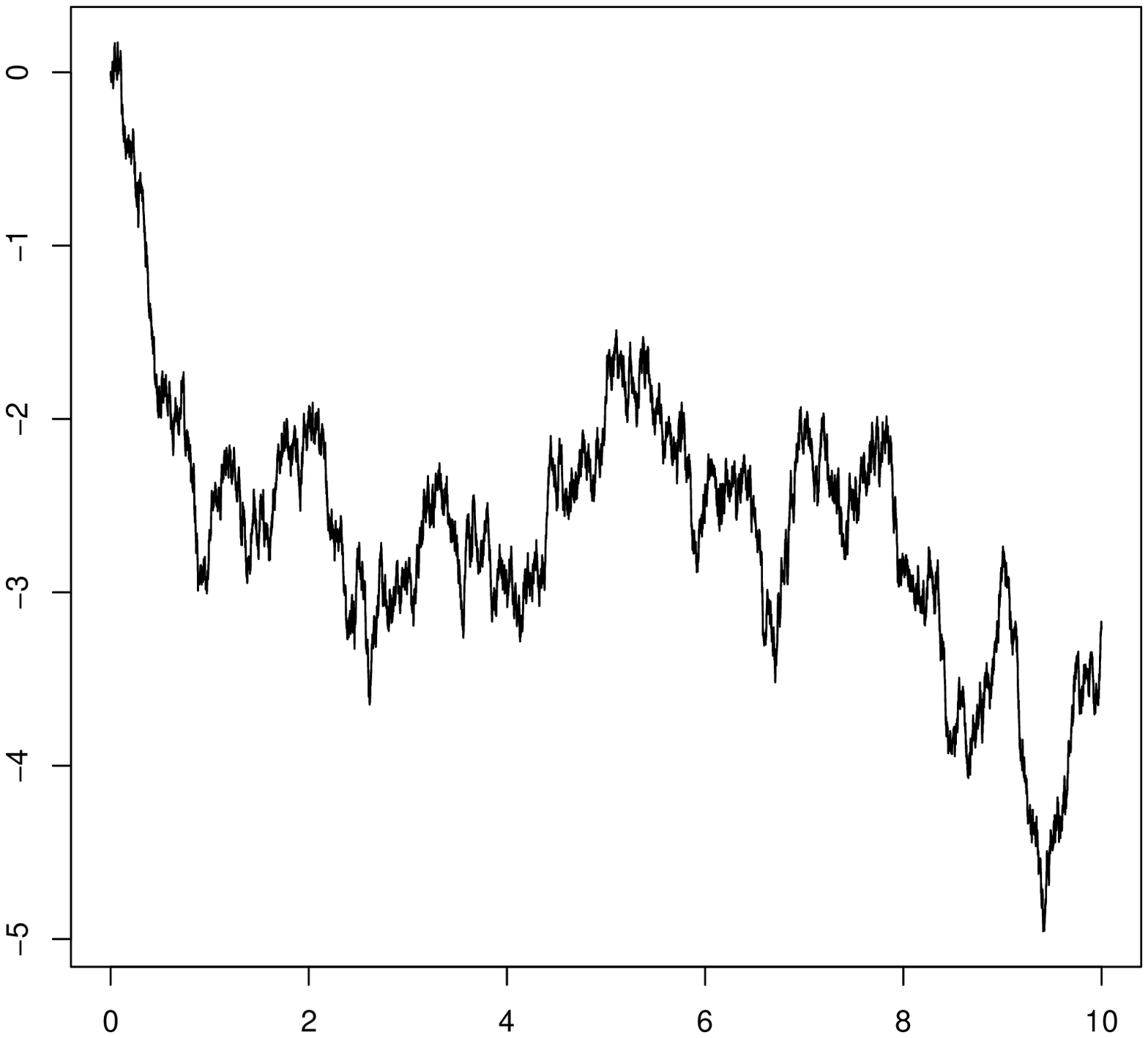}  
\includegraphics{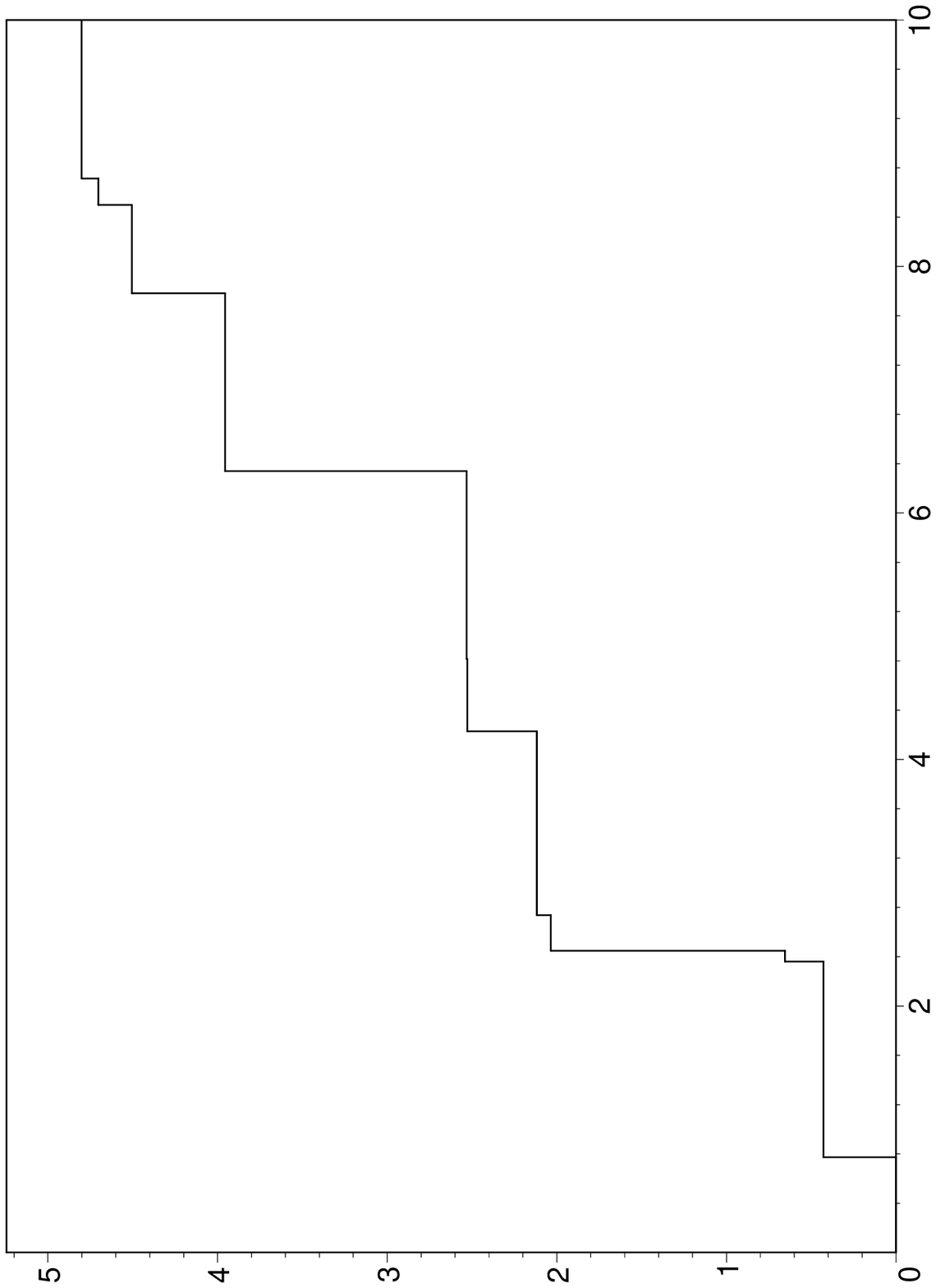} 
\caption{$W(x)$ against $x$ for a particular realisation of (a) a standard Brownian motion and (b) 
a driftless compound Poisson process with exponentially-distributed jumps.}
\label{interlacingFigure} 
\end{figure}

Roughly speaking, the paths of a L\'{e}vy process consist of intervals of drifted Brownian motion separated by jumps whose height and frequency are controlled by
the L\'{e}vy measure.
In the particular case where the L\'{e}vy measure ${m}$ is {\em finite}, i.e.
\begin{equation}
  \rho := \int_{\mathbb R} {m} ( \d y) < \infty
  \label{finiteLevyMeasure}
\end{equation}
we call $W$ an {\em interlacing process}; it may be expressed in the form
\begin{equation}
W (x) = 2a\, x + 2\sqrt{g}\, B(x) + J(x)
\label{interlacingProcess}
\end{equation}
where $B$ is a standard Brownian motion, the {\em drift} $2a$ is given by
\begin{equation}
  2 a := 2 \mu g - \int_{\mathbb R} \frac{y}{1+y^2} \,{m} (\d y)
  \label{drift}
\end{equation}
and
\begin{equation}
  J(x) = \sum_{j=1}^{P(x)} h_j
\label{jumps}
\end{equation}
where $P$ is a Poisson process of intensity $\rho$, and the $h_j$ are random variables with probability measure ${m}/\rho$.
The processes $B$ and $P$, and the random variables $h_j$, in these expressions are mutually independent. Figure \ref{interlacingFigure}
displays realisations of two important particular cases: 
(a) a standard Brownian motion, corresponding to the choice 
$a=\rho=0$,
and 
(b) a compound Poisson process with exponentially-distributed jumps $h_j$, corresponding to the choice $a=g=0$ and $\rho = 1$.

\begin{figure}[htbp]
\vspace{7cm}  
\includegraphics{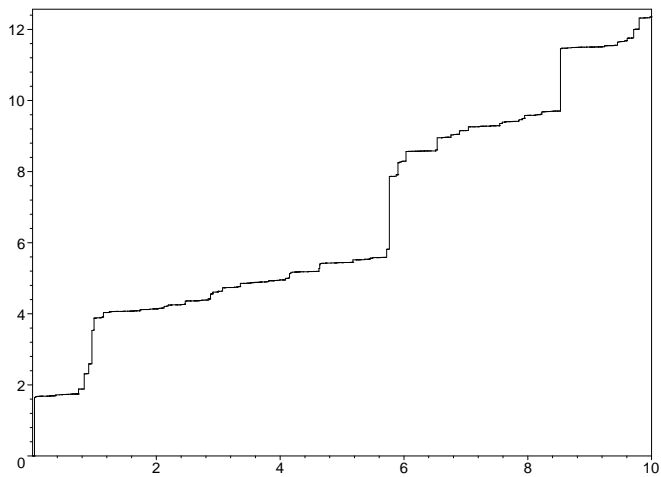} 
\begin{picture}(0,0)
\end{picture}
\caption{$W(x)$ against $x$ for a particular realisation of the subordinator 
such that $a=0$ and the L\'{e}vy measure is given by
Equation (\ref{hermiteLevyMeasure}) with $q=1$.} 
\label{nonInterlacingFigure} 
\end{figure}

Not every L\'{e}vy measure is finite, and so interlacing processes do not exhaust the class of L\'{e}vy processes; see 
Figure \ref{nonInterlacingFigure} for a particular realisation of a L\'{e}vy process that is {\em not} interlacing, and \cite{Ap2}
for a description of other important examples. Nevertheless, every
L\'{e}vy process $W$ may be approximated by an interlacing process $W_\varepsilon$ as follows: let $\varepsilon >0$, and define
from ${m}$ a new L\'{e}vy measure ${m}_\varepsilon$ by
$$
{m}_\varepsilon ( A ) := {m} ( A_\varepsilon ) \;\; \text{where} \;\; A_\varepsilon := \left \{ y \in A \, : \; |y| > \varepsilon \right \}\,.
$$
The measure ${m}_\varepsilon$ is finite. Hence the process $W_\varepsilon$ with L\'{e}vy characteristics $\mu g$, $g$ and ${m}_\varepsilon$ is an interlacing process, and
$\Lambda_\varepsilon$ converges to $\Lambda$ pointwise as $\varepsilon \rightarrow 0$. The sense in which the paths of $W_\varepsilon$ approximate those of $W$ can be made precise;
see the proof of Theorem 1 in \cite{Be}. The intuitive picture suggested by this construction is that
non-interlacing processes experience very small jumps with a very high frequency, as illustrated by Figure \ref{nonInterlacingFigure}.

In what follows, particular attention will be paid to the {\em subordinator} case, where the L\'{e}vy process is non-decreasing.
For such processes, $g=0$ and the L\'{e}vy measure does not charge $(-\infty,0]$. A subordinator is therefore of bounded variation on finite intervals and there holds
$$
\int_0^\infty \min \{1,y\} \,{m} (\d y) < \infty
$$
and
\begin{equation}
  \Lambda ( \theta) = 2 a
  + \int_0^\infty \left ( \e^{\i y \theta} - 1 \right ) \,{m} (\d y)
\label{levyExponentForSubordinators}
\end{equation}
with $a\ge 0$. 
We shall see that this case, as in Kotani's study, affords a number of simplifications.

\subsection{The Frisch--Lloyd--Kotani study}
\label{KotaniSubsection}
Let us review the formal steps involved in the calculation of $\Omega$ when the potential is the distributional derivative of a L\'{e}vy process $W$. We refer the reader to the original papers
\cite{FL,Ko} for detailed explanations, hypotheses and proofs, and to \cite{CTT} for heuristics.

Written in terms of the Riccati variable
$$
Z(x) = \frac{\psi'(x,E)}{\psi(x,E)}
$$
the Schr\"{o}dinger equation (\ref{schroedingerEquation}) with the potential $V=W'=w$ becomes
$$
Z' = - E - Z^2 + w\,.
$$
$Z$ is a Feller process, started at infinity, whose infinitesimal generator, say ${\mathscr G}$, is given by (see for instance \cite{Ap1,RY})
$$
  \left ( {\mathscr G} u \right ) (z) = \left ( 2\mu g - E - z^2 \right )
  u'(z) + 2g\, u''(z) + \int_{{\mathbb R}} \left [
    u(z+y)-u(z)- \frac{y\, u'(z)}{1+y^2} \right ]\,{m}(\d y)\,.
$$
$Z$ has a stationary distribution whose probability density, denoted $f$, is in the kernel of the adjoint ${\mathscr G}^\dag$. Hence
$$
\left ( E - 2\mu g + z^2 \right ) f(z) + 2g\, f'(z) + \int_{\mathbb R} \left [ \int_z^{z-y} f(t)\,\d t + \frac{y\, f(z)}{1+y^2} \right ]\,{m}(\d y) = N
$$
where the integrated density of states appears as a constant of integration; $N$ may be determined by finding a positive integrable solution
and imposing the normalisation condition. The Lyapunov exponent $\gamma$ is then given by the Cauchy Principal Value integral
$$
\gamma = \dashint_{-\infty}^\infty z f(z)\,\d z\,.
$$
The density $f$ contains more information than is strictly necessary to determine $\Omega$, and it is sometimes more convenient to work with its Fourier transform 
$$
\check{f}(s)
= \int_{\mathbb R} \e^{-\i s z} f(z) \,\d z\,.
$$
$\check{f}$ satisfies the differential equation
\begin{equation}
\check{f}''(s) + c(\i s)  \check{f} (s) = E \check{f}(s)\,, \;\; s \ne 0\,,
\label{differentialEquation}
\end{equation}
where
\begin{equation}
  c(s) := -\frac{\Lambda(\i s)}{s}
  = 2g [ \mu - s ]  + 
  \int_{{\mathbb R}} 
   \left[ \frac{1- \e^{-s y}}{s} - \frac{y}{1+y^2} \right ] \, 
   m( \d y)
  \,.
  \label{coefficient}
\end{equation}
It is then readily verified (see for instance \cite{Ha}, Appendix B) that
\begin{equation}
  \Omega = \i \frac{\check{f}'(0+)}{\check{f}(0+)}
\label{frischLloydFormula}
\end{equation}
where, with a slight abuse of notation, $\check{f}$ now denotes {\em any} non-trivial solution of the differential equation (\ref{differentialEquation})
such that
$$
\lim_{s \rightarrow \infty} \check{f}(s) = 0\,.
$$
This representation in terms of the decaying solution of a
homogeneous second-order linear differential equation was used by
Frisch \& Lloyd as the basis of their numerical study of the density of states \cite{FL}. Explicit formulae 
were obtained by Halperin in the case of a Brownian motion  \cite{Ha} and by Nieuwenhuizen in the case of a compound Poisson process with a gamma distribution of the jumps
\cite{Ni}.
Kotani performed a semiclassical analysis of the differential equation, in the limit of low energy, for the particular case where $W$ is a subordinator \cite{Ko}.
By using Langer's transformation, he was able to obtain an approximation of the decaying solution valid uniformly in the variable $s$, from 
which the low-energy behaviour of $N(E)$ could be deduced.

\subsection{The supersymmetric case}
\label{supersymmetricSubsection}
The Schr\"{o}dinger
equation with the supersymmetric potential (\ref{supersymmetricPotential}) can be
recast as the Dirac system 
\begin{align}
\label{dirac1}
- \psi' - \frac{w}{2} \,\psi &= \sqrt{E} \,\phi \\
\label{dirac2}
  \phi' -  \frac{w}{2} \,\phi &= \sqrt{E} \,\psi \,.
\end{align}
The meaning of these equations when $w=W'$ is a L\'{e}vy noise becomes clear if we introduce an
integrating factor:
\begin{align}
\label{integratedDirac1}
- \frac{\d}{\d x} \left [ \exp \left ( \frac{W}{2} \right ) \psi \right ] &= \sqrt{E} \exp \left (  \frac{W}{2} \right ) \phi \\
\label{integratedDirac2}
\frac{\d}{\d x} \left [ \exp \left (   - \frac{W}{2} \right ) \phi \right ] &= \sqrt{E} \exp \left ( - \frac{W}{2} \right ) \psi\,.
\end{align}

Let us outline the modifications of the Frisch--Lloyd--Kotani approach required in the supersymmetric case. The appropriate Riccati variable is
\begin{equation}
Z := -\sqrt{E} \,\frac{\phi}{\psi}\,.
\label{riccatiVariable}
\end{equation}
Then
\begin{equation}
Z' = - E - Z^2 + w Z
\label{riccatiEquation}
\end{equation}
and the L\'{e}vy noise now appears as a {\em multiplicative} term; as we have written it,
this stochastic equation should be interpreted in the sense
of Stratonovich. 
In the supersymmetric case, the essential spectrum is contained in ${\mathbb R}_+$.
It will be convenient to consider in the first instance the case $E < 0$, so that $N(E)=0$ and $Z$ has a stationary distribution supported in ${\mathbb R}_+$, whose  density we denote again
by $f$. To find an equation for $f$, we start from the simple observation that the {\em logarithm} of $Z$ satisfies a stochastic equation in which
the noise appears as an {\em additive} term, just as in the previous subsection. It is then straightforward to deduce
\begin{multline}
  \left ( E - 2\mu g\, z + z^2 \right ) f(z)  
 + 2g\, z \frac{\d}{\d z} \left [ z f(z) \right ]  \\
  + \int_{\mathbb R} \left \{ 
   \int_z^{z \e^{-y}} f (t)\,\d t + \frac{y z}{1+y^2} f(z) 
  \right \} {m} (\d y) = N\,.
\label{frischLloydEquation}
\end{multline}

\subsection{Exponentials of L\'{e}vy processes}
\label{exponentialsSubsection} 
For $E=0$, Equation (\ref{riccatiEquation}) has a non-negative solution given by
$$
\frac{1}{Z(x)} = \e^{-W(x)} \frac{1}{Z(0)} + \int_{0}^x \e^{-\left [ W(x)-W(t) \right ]}\, \d t\,.
$$
In particular, if $Z(0)=\infty$, we may use the stationarity of the increments of $W$ to deduce that
$$
\frac{1}{Z(x)}  \eqlaw
\int_{0}^x \e^{-W(t)}\, \d t\,.
$$
So the reciprocal of the zero-energy Riccati variable has the law of an exponential of the L\'{e}vy process $W$. 
The study of these exponentials has received a great deal of attention in the probability literature \cite{BY}, and its close connection with
our disordered system makes it a rich source of ideas. In particular,
Bertoin \& Yor \cite{BY2} studied the {\em moments} of the exponentials and found for them a first-order recurrence relation
which we proceed to extend to the case $E \ne 0$.

\subsection{Main results and outline of the paper}
\label{outlineSubsection}

In the supersymmetric case, because the noise appears in 
Equation (\ref{riccatiEquation}) as a multiplicative term, 
the equation for the Fourier transform of $f$ contains an awkward integral 
term. 
Following the example of Bertoin \& Yor, 
it is in fact more expedient to work with the {\em Mellin transform}
\begin{equation}
  \hat{f}(s) := \int_{0}^{\infty} z^{-s} f(z)\,\d z\,.
  \label{mellinTransform}
\end{equation}
We derive in \S \ref{frischLloydSection} the supersymmetric counterparts
of the Frisch--Lloyd formulae (\ref{differentialEquation}) and (\ref{frischLloydFormula}):
for $E<0$,
\begin{equation}
  E \hat{f}(s+1) - c(s)  \hat{f}(s) + \hat{f}(s-1) = 0
  \label{differenceEquation}
\end{equation}
and
\begin{equation}
  \Omega (E) = \frac{c(0)}{2} - E \frac{\hat{f}(1)}{\hat{f}(0)}\,.
  \label{mellinFormula}
\end{equation}
The coefficient $c(s)$ appearing in the {\em difference equation} (\ref{differenceEquation}) was defined earlier by Equation (\ref{coefficient})
for $s \ne 0$, and this definition may be extended to $s=0$ by taking the obvious limit. 
To find $\hat{f}(0)$ and $\hat{f}(1)$, one can either solve the integro-differential equation \eqref{frischLloydEquation} for the distribution and then compute the relevant integrals or, alternatively, seek a positive solution of 
Equation \eqref{differenceEquation} satisfying the normalisation condition $\hat{f}(0)=1$.
The problem of computing the density of states
and the inverse localisation length in the ``physical'' region $E>0$ is thus reduced to a problem of analytic continuation---
in the energy variable---
from ${\mathbb R}_-$ to the cut plane ${\mathbb C} \backslash {\mathbb R}_+$.
This formulation in terms of the Mellin transform sheds new light on the few 
solvable cases that have appeared in the literature; we review them in \S \ref{reviewSection}.

The remaining sections are devoted to the subordinator case, where we have an effective tool for performing the analytic continuation. We show
in \S \ref{continuedFractionSection} that
\begin{equation}
  \Omega (E) = \frac{c(0)}{2} + \cfrac{-E}{c(1) + \cfrac{-E}{c(2) + \cdots}}
  \label{continuedFraction}
\end{equation}
for every complex $E$ outside the essential spectrum, and we remark that the calculation of the density of states is tantamount to solving a Stieltjes
moment problem. In \S \ref{hermiteSection}, we look for instances where the general solution of the difference equation (\ref{differenceEquation}) is known
explicitly. For the subordinator with L\'{e}vy measure
\begin{equation}
  m(\d y) = \frac{2 p }{\sqrt{\pi}} 
  \,\frac{\d}{\d y} 
  \left[ \frac{-\e^{-q y}}{\sqrt{1-\e^{-2y}}} \right]\,\d y\,, \quad p,\,q > 0,
\label{hermiteLevyMeasure}
\end{equation}
we find, with the help of Masson \cite{Ma}, that the complex Lyapunov exponent is expressible in terms of the parabolic cylinder functions.
Then, in \S \ref{semiclassicalSection}, we take up the problem of determining the low-energy behaviour of the integrated density of states and show how to compute the leading term in a semiclassical approximation;
we also provide a list of the possible singular behaviours.
Some of the implications of these results for diffusions in a random environment are discussed briefly.
We end the paper in \S \ref{conclusionSection} with a few concluding remarks.

\section{Characteristic function and Mellin transform}
\label{frischLloydSection}
This section provides a derivation of Equations (\ref{differenceEquation}) and (\ref{mellinFormula}). For the
particular case where $W$ is a subordinator with finite means and a finite L\'{e}vy measure, this derivation may be made completely rigorous by adapting
the arguments of Frisch \& Lloyd \cite{FL} and Kotani \cite{Ko}. In the general case, however, we shall be content to
view the formulae as merely plausible.

\subsection{The difference equation}
\label{differenceEquationSubsection}

First, as a consequence of Sato's Theorem 25.3 \cite{Sa}, 
we have the following criterion  for the {\em existence} of the coefficient $c(s)$ defined by Equation (\ref{coefficient}):
for $s \ne 0$, $c(s)$ exists if and only if
$$
\int_{|y|>1} \text{\em e}^{-s y}\, {m}( \text{\em d} y) < \infty\,.
$$
Also, it is readily seen by direct calculation that $c(0)$ exists if and only if
\begin{equation}
\int_{\mathbb R} y\, {m} (\d y) < \infty
\label{subordinatorJumpCondition}
\end{equation}
and, in case of existence,
$$
c(0) = {\mathbb E} \left ( W(1) \right )\,.
$$
In particular, $c(s)$ exists for every $s \ge 0$ if $W$ is a subordinator with finite means.

Next, to derive the difference equation, multiply Equation (\ref{frischLloydEquation}) by $z^{-s}$ and integrate over $z$. We use
\begin{equation}
\notag
\int_0^{\infty} z^{-s}\, z \frac{\d}{\d z} \left [ z f(z) \right ] \,\d z
= z^{1-s} z f(z) \Bigl |_0^{\infty} - (1-s) \int_0^{\infty} z^{-s}\, z f(z) \,\d z 
= (s-1) \hat{f}(s-1)
\end{equation}
where we have assumed that
\begin{equation}
\lim_{z \rightarrow 0+} z^{2-s} f(z) = 0\;\;\text{and}\;\;\lim_{z \rightarrow \infty} z^{2-s} f(z) = 0\,.
\label{tailCondition}
\end{equation}
Also,
\begin{multline}
\notag
\int_0^{\infty}  z^{-s} \left [  \int_z^{z \e^{-y}} f (t)\,\d t + \frac{y z}{1+y^2} f(z) \right ] \d z \\
= \int_0^{\infty}  z^{-s} \left [  \int_z^{z \e^{-y}} f (t)\,\d t \right ] \,\d z + \frac{y}{1+y^2} \hat{f}(s-1) \\
= \frac{1}{1-s} z^{1-s} \left [  \int_z^{z \e^{-y}} f (t)\,\d t \right ] \Bigl |_0^{\infty} \\ 
- \frac{1}{1-s} \int_0^{\infty} z^{1-s} 
\left [  f (z \e^{-y} ) \, \e^{- y} - f(z) \right ]\,\d z + \frac{y}{1+y^2} \hat{f}(s-1) \\
= \frac{1}{s-1} \int_0^{\infty} z^{1-s} 
f (z \e^{-y} ) \, \e^{- y} \,\d z  - \frac{1}{s-1} \hat{f}(s-1) + \frac{y}{1+y^2} \hat{f}(s-1)\,.
\end{multline}
By making the obvious substitution in the integral, we obtain 
\begin{multline}
\notag
\int_0^{\infty}  z^{-s} \left [  \int_z^{z \e^{-y}} f (t)\,\d t + \frac{y z}{1+y^2} f(z) \right ] \d z \\
= \frac{1}{s-1} \left [ \e^{-(s-1) y} - 1 + \frac{(s-1) y}{1+y^2} \right ] \hat{f}(s-1)\,.
\end{multline}
When we put these results together, we eventually find
that $\hat{f}(s)$ solves the difference equation (\ref{differenceEquation}).

Let us now look back on Condition (\ref{tailCondition}) in the case $s \ge 0$. The fact that the limit at infinity vanishes
is a consequence of the Rice formula
$$
\lim_{z \rightarrow \infty} z^2 f(z) = N = 0\,.
$$
That the limit at zero vanishes is obvious if $W$ is a subordinator and $E=-k^2<0$, because the support of $f(z)$
is contained in $[k,\infty)$.

\subsection{The Lyapunov exponent}
\label{lyapunovSubsection}
For $E \ne 0$,
\begin{equation}
\notag
\frac{1}{x} \ln \left | \psi(x) \right | = \frac{1}{x} \ln  \left | \sqrt{E} \frac{\phi(x)}{Z(x)} \right |
= \frac{1}{x} \ln \sqrt{| E |} + \frac{1}{x} \ln \left | \phi(x) \right | - \frac{1}{x} \ln \left | Z(x) \right |\,.
\end{equation}
If the L\'{e}vy process is not a pure drift then the Riccati process will behave ergodically, and so  
the alternative formula
\begin{equation}
\gamma (E) = \lim_{x \rightarrow \infty} \frac{1}{x} \ln \left | \phi (x) \right |
\label{lyapunovInTermsOfPhi}
\end{equation}
will hold for $E \ne 0$.
Our aim is to use this formula to express the Lyapunov exponent in terms of the Mellin transform
of the Riccati variable. 

We shall suppose that
$W$ is an interlacing process of the form (\ref{interlacingProcess}-\ref{jumps}), with a finite mean at $x=1$. By Equation (\ref{subordinatorJumpCondition})
this implies that the jumps $h_j$ themselves have a finite mean.
The Dirac system (\ref{dirac1}-\ref{dirac2}) is then
\begin{align*}
  - \psi' 
  - \bigg[ 
     a + \sqrt{g}\, B' + \frac12\sum_{j=1}^\infty h_j \delta_{x_j} 
     \bigg]
  \,\psi &= \sqrt{E} \,\phi \\
  \phi' 
  -\bigg[ 
     a + \sqrt{g}\, B' + \frac12\sum_{j=1}^\infty h_j \delta_{x_j} 
     \bigg] \,\phi &= \sqrt{E} \,\psi\,.
\end{align*}
Here
$$
x_0 := 0 \quad \text{and} \quad x_{j+1} := x_j + \ell_j
$$
where the $\ell_j$ are independent and exponentially distributed with parameter $\rho$.  For $x \in (x_j,\,x_{j+1})$, we have
\begin{align*}
- \psi' - \left[ a + \sqrt{g}\, B' \right] \,\psi &= \sqrt{E} \,\phi \\
  \phi' - \left[ a + \sqrt{g}\, B' \right] \,\phi &= \sqrt{E} \,\psi\,.
\end{align*}
Hence
\begin{multline}
\notag
\ln | \phi (x_{j+1}-) | = \ln | \phi (x_j+) | + \int_{x_j}^{x_{j+1}} \frac{\phi'(y)}{\phi (y)}\, \d y \\
= \ln | \phi (x_j+) | 
  + a\ell_j + \sqrt{g}\, \left[ B(x_{j+1}) - B(x_j) \right]  
  + \int_{x_j}^{x_{j+1}} \frac{-E}{Z(y)} \,\d y\,.
\end{multline}
On the other hand, by making use of Equation (\ref{integratedDirac2}), we find
$$
\phi (x_j+)  = \exp \left ( \frac{h_j}{2} \right )  \phi (x_j-)
$$
and it follows that
$$
\ln | \phi (x_{j+1}-) | = \ln | \phi (x_j-) | 
+ a\ell_j + \sqrt{g}\, \left[ B(x_{j+1}) - B(x_j) \right] 
+ \frac{h_{j+1}}{2}
+ \int_{x_j}^{x_{j+1}} \frac{-E}{Z(y)}\,\d y\,.
$$
Then, by summing over $j$, we obtain
\begin{equation}
\notag
\frac{1}{x} \ln | \phi (x) | = \frac{1}{x} \ln | \phi (0) | + \frac{1}{x} 
\left \{
a x + \sqrt{g}\, B(x) +  \frac12\sum_{j=1}^{P(x)} h_j \right \} + \frac{1}{x} \int_0^x \frac{-E}{Z(y)} \,\d y\,.
\end{equation}
In the limit as $x \rightarrow \infty$, this becomes
\begin{equation}
\gamma = \frac{c(0)}{2}  +  \int_{0}^\infty  \frac{-E}{z} f(z)\,\d z\,.
\label{lyapunovAscending}
\end{equation}
This establishes the formula (\ref{mellinFormula}) for the complex Lyapunov exponent when $W$ is an interlacing process with finite means. 
We expect this formula to hold more generally even in cases where the L\'{e}vy measure is not finite since, as
mentioned in \S \ref{levySubsection}, every such measure may be approximated by a finite measure.

\section{Some known solvable cases}
\label{reviewSection}
There are very few known cases where the complex Lyapunov exponent may be expressed in terms of familiar
functions. Whereas the examples we review here were discovered by solving the integro-differential equation (\ref{frischLloydEquation}) directly, our
presentation will emphasise the alternative approach based on solving the difference equation satisfied by the Mellin transform. 

\begin{example}
When the L\'{e}vy process is a pure drift, the process is deterministic and we have
$$
\Lambda (\theta) = 2\i a\, \theta \;\; \text{and} \;\; c(s) = 2 a\,.
$$
The difference equation (\ref{differenceEquation}) takes the simple form
\begin{equation}
E \hat{f}(s+1) - 2a \hat{f}(s) + \hat{f}(s-1) = 0\,.
\label{pureDriftDifferenceEquation}
\end{equation}
Its general solution is
$$
   c_+ \left ( a + \sqrt{a^2- E} \right )^{-s} 
 + c_- \left ( a - \sqrt{a^2-E} \right )^{-s}\,.
$$
The Mellin transform of the invariant density is positive for every $s$ and equals unity for $s=0$. Hence
$$
\hat{f}(s) = \left ( a+\sqrt{a^2-E} \right )^{-s}
$$
and, by using Equation (\ref{lyapunovAscending}), we find
\begin{equation}
\Omega =  \sqrt{a^2-E}\,.
\label{characteristicFunctionForPureDrift}
\end{equation}
The spectrum is the interval $E \ge a^2$; there, we have
$$
N(E) = 
\frac{1}{ \pi} \sqrt{E-a^2} \;\;\text{and}\;\; \gamma (E) = 0\,.
$$
Hence $\gamma$ vanishes in the spectrum--- reflecting the fact that, 
for the non random potential $V(x)=a^2$,
there is no localisation.
\label{pureDriftExample}
\end{example}

\begin{example}
The case where the L\'{e}vy process is a Brownian motion with drift has been studied independently by Ovchinnikov and Erikhman \cite{OE}
and Bouchaud {\em et al.} \cite{BCGL}.
In this case,
$$
\Lambda (\theta) = 2g \left [ \i \mu \theta - \theta^2 \right ]
  \;\; \text{and} \;\; c(s) = 2g\, [\mu- s]\,.
$$
Hence $\hat{f}(s)$ satisfies the difference equation
$$
E \hat{f} (s+1) +2g \, [ s-\mu ]\,\hat{f}(s) + \hat{f}(s-1) = 0
\:.
$$
We recognise the reccurence relation satisfied by the modified Bessel functions.
The general solution is
$$
  \left [ \sqrt{-E} \right ]^{-s} 
  \left\{ c_K\,  K_{s-\mu} \left( \sqrt{-E}/g \right) 
 + (-1)^{s-\mu} c_I\, I_{s-\mu} \left( \sqrt{-E}/g \right) \right \}
$$
and we obtain a positive solution by taking $c_I=0$. 

Alternatively, we can solve the
equation \eqref{frischLloydEquation} for the invariant density. Since, in this case, the L\'{e}vy measure is identically zero, this equation reduces
to a differential equation. The relevant solution is
(before normalisation) 
$$
  f(z) = z^{\mu-1} \exp \left[ -\frac{1}{2g} 
  \left( z - \frac{E}{z} \right) \right] {\mathbf 1}_{(0,\infty)} (z)
$$
and its Mellin transform 
$$
\hat{f}(s) = 2 \left [ \sqrt{-E} \right ]^{\mu-s} 
K_{s-\mu} \left( \sqrt{-E}/g \right )
$$
agrees, as it should, with the expression found earlier by solving the
difference equation.
Hence, by Formula (\ref{mellinTransform}),
$$
\Omega (-k^2) = \mu g 
  + k \frac{ K_{1-\mu} \left( k/g \right)}{K_{-\mu} \left ( k/g \right )}
$$
where we have set $E=-k^2$. The analytic continuation of $\Omega$ from ${\mathbb R}_-$ to ${\mathbb R}_+$ may be done ``by hand'' and
consists of replacing $k$ by $-\i k$:
$$
\Omega (k^2+\i 0+) = \mu g + k 
\frac{H_{1-\mu}^{(1)} \left( k/g \right)}{H_{-\mu}^{(1)} \left( k/g \right )}\,.
$$
Expressions for $N$ and $\gamma$ follow easily.
In particular, for $\mu=0$ (the driftless case), we find
$$
N(E) \sim \frac{2g}{\ln^2 E} \;\;\text{and}\;\; 
\frac{1}{\gamma (E)} \sim -\frac{1}{2g} \ln E \quad \text{as $E \rightarrow 0+$}\,.
$$
So the density of states has a so-called Dyson-type singularity at the bottom edge of the spectrum, and the model
exhibits a very interesting ``delocalisation'' phenomenon there
\cite{BCGL,CT,TexHag10}.

\label{brownianWithDriftExample}
\end{example}

\begin{example}
Let the L\'{e}vy exponent be
$$
  \Lambda (\theta) 
  = \rho \int_0^{\infty} 
  \left[ \e^{\i \theta y}-1\right]\,q \e^{-q y} \,\d y 
  = \rho \frac{\i \theta}{q-\i \theta}\,.
$$
This corresponds to a driftless compound Poisson process of intensity $\rho$ where the jumps are exponentially distributed with parameter $q$. Then
$$
c(s) = \frac{\rho}{q+s}\,, \quad s \ge 0\,.
$$
Set
$$
E = -k^2, \quad k > 0\,.
$$
Masson \cite{Ma} showed that the general solution of the difference equation (\ref{differenceEquation}) is given by
\begin{equation}
\notag
c_+ u_+(s) + c_- u_-(s)
\end{equation}
where
\begin{multline}
  \notag
  u_{\pm} (s) := \left ( \pm k \right )^{-s} 
  \text{\tt B} \left ( \pm \frac{\rho}{2 k}, q+s+1 \right ) \\
  \times {_2}F_1\!\left ( 
     q+s+1, \pm \frac{\rho}{2 k}+1;q+s+1 \pm \frac{\rho}{2 k};-1 
  \right )
\end{multline}
and $\text{\tt B}$ is the Euler beta function.
To obtain a positive solution, we set $c_-=0$. The formula
eqref{mellinFormula} for the complex Lyapunov exponent then agrees
with the calculation of 
our previous article \cite{CTT}, where we have
worked directly from the  
invariant density (before normalisation), namely
$$
f(z) = \frac{z^{-q}}{z^2-k^2} \left ( \frac{z-k}{z+k} \right )^{\frac{\rho}{2 k}}
\,{\mathbf 1}_{\left ( k,\infty \right )}(z)
\,.
$$
After normalisation we obtain the Lyapunov exponent from
Equation \eqref{lyapunovAscending} \cite{CTT}. 
For negative energies $\gamma$ coincides with the characteristic 
function $\Omega$.
After analytic continuation we find the integrated density of states. 
The low-energy behaviour is (see Bienaim\'{e} \cite{Bi})
\begin{equation}
N(E) \sim  \frac{\rho^{2q+1}}{\Gamma(q+1)^2} \,E^{-q} \,\exp \left [ -\frac{\pi}{2}\frac{\rho}{\sqrt{E}} \right ] \quad \text{as $E \rightarrow 0+$}
\label{bienaime}
\end{equation}
and
$$
\gamma\sim\frac{\rho}{2q} \quad \text{as $E \rightarrow 0+$}\,.
$$
\begin{remark}
The low-energy behaviour of the density of states in the supersymmetric case is quite different from the usual Lifshits behaviour \cite{LGP}, namely
$$
N(E) \sim \exp \left [-\pi \frac{\rho}{\sqrt{E}} \right ] \quad \text{as $E \rightarrow 0+$}\,.
$$
The physical reasons for this difference will be explained in a
forthcoming paper. 
\label{lifshitsRemark}
\end{remark}
\label{exponentiallyDistributedJumpsExample}
\end{example}

\section{Continued fractions}
\label{continuedFractionSection}

There is a well-known connection between linear second-order
difference equations and continued fractions \cite{Ga,Ma}; we exploit it to derive
a continued fraction that may, at least in the subordinator case, be used to compute the complex Lyapunov exponent.

\subsection{Pincherle's Theorem}

We say that a non-zero solution $u_n$ of the difference equation
\begin{equation}
u_{n+1} = a_n u_{n-1} + b_n u_n \,, \quad n \in {\mathbb N}\,,
\label{integerDifferenceEquation}
\end{equation} 
is {\em recessive} (or minimal or subdominant) if, for every other linearly independent solution $v_n$, there holds
$$
u_n = o \left ( v_n \right ) \;\; \text{as $n \rightarrow \infty$}\,.
$$
Pincherle's Theorem says that
the continued fraction
$$
K := \cfrac{a_1}{b_1 + \cfrac{a_2}{b_2 + \cdots}}
$$
converges if and only if the difference equation (\ref{integerDifferenceEquation}) has a recessive solution $u_n$ with $u_0 \ne 0$. In case of convergence,
the limit is $-u_1/u_0$ where $u_n$ is any such recessive solution.
The proof of this well-known result is elementary, but we include it here because it contains information that will be useful later on.
Introduce two particular solutions $p_n$ and $q_n$ of the difference equation (\ref{integerDifferenceEquation}) satisfying
$$
p_0 = q_1 = 1 \;\;\text{and}\;\; q_0=p_1=0\,.
$$
These solutions are obviously linearly independent, and every other solution may be expressed as a linear combination of them.
Set
$$
K_n := \frac{p_n}{q_n}\,,\;\; n \in {\mathbb N}\,.
$$
To say that the continued fraction converges is to say that $K_n$ has a (finite) limit as $n \rightarrow \infty$.
First, suppose that the continued fraction converges; call its limit $K_{\infty}$ and set
$$
u_n := q_n K_{\infty} - p_n\,.
$$
It then follows from this definition of the sequence $u_n$ that $u_0=-1$ and
$$
K_{\infty} = -\frac{u_1}{u_0}\,.
$$
Furthermore, $u_n$ solves the difference equation. To show that $u_n$ is recessive, suppose that $v_n$ is any other linearly independent solution. 
We can express $v_n$ in the form
$$
v_n = \alpha u_n + \beta q_n
$$
for some number $\alpha$ and some non-zero number $\beta$. Then
$$
\frac{v_n}{u_n} = \alpha + \beta \frac{q_n}{u_n} = \alpha + \frac{\beta}{K_{\infty}-\frac{p_n}{q_n}} \xrightarrow[n \rightarrow \infty]{}  \beta \times \infty\,.
$$
This shows that $u_n$ is recessive. 
Conversely, suppose that the difference equation has a recessive solution $u_n$ such that $u_0 \ne 0$. Since $q_0=0$,
the particular solutions $q_n$ and $u_n$ are linearly independent. Therefore
$$
\lim_{n \rightarrow \infty} \frac{u_n}{q_n} = 0\,.
$$
On the other hand, we can express $u_n$ as a linear combination of $p_n$ and $q_n$:
$$
u_n = u_0 p_n + u_1 q_n\,.
$$
It follows that
$$
\frac{p_n}{q_n} = \frac{1}{u_0} \left ( \frac{u_n}{q_n} - u_1 \right ) \xrightarrow[n \rightarrow \infty]{} - \frac{u_1}{u_0}\,.
$$
To complete the proof, there only remains to observe that there cannot be {\em two} linearly independent solutions that are recessive.

\begin{remark}
\label{recessiveRemark}
The proof shows in particular that, if $K$ converges to $K_{\infty}$, then every recessive solution of the difference equation is a multiple of
$$
q_n K_{\infty}-p_n\,.
$$
\end{remark}

\subsection{The subordinator case}
\label{subordinatorSubsection}

Let us now elaborate the relevance of Pincherle's Theorem to our study. For convenience, we shall omit the drift, so that the L\'{e}vy exponent is given by
Equation (\ref{levyExponentForSubordinators}) with $a=0$, and the coefficient $c(n)$ by
\begin{equation}
c(n) = \int_0^{\infty} \frac{1-\e^{-n y}}{n}\, {m} (\d y)\,.
\label{subordinatorCoefficient}
\end{equation}
Set
\begin{equation}
u_n := E^n \hat{f}(n)\,, \; n \in {\mathbb N}\,.
\label{mellinTransformAsAscendingSolution}
\end{equation}
The fact that $\hat{f}(s)$ solves the difference equation (\ref{differenceEquation}) for $s \in {\mathbb N}$ implies that $u_n$ solves 
\begin{equation}
u_{n+1} = -E \,u_{n-1} + c(n) \,u_n\,, \quad n \in {\mathbb N}\,.
\label{ascendingDifferenceEquation}
\end{equation}
The corresponding continued fraction is
\begin{equation}
K(E) := \cfrac{-E}{c(1) + \cfrac{-E}{c(2) + \cfrac{-E}{c(3) + \cdots}}}\,.
\label{ascendingContinuedFraction}
\end{equation}

We now observe that the coefficients $c(n)$ are
obviously {\em positive}. 
This means 
that the continued fraction
(\ref{ascendingContinuedFraction}) is of the kind studied by Stieltjes \cite{BO,NS,St,Wa}. 
Stieltjes showed that a 
necessary and sufficient condition for the convergence of the continued fraction in ${\mathbb C} \backslash {\mathbb R}_+$ is that the series
$$
\sum_{n=1}^\infty c(n)
$$
diverge. Now, by Equation (\ref{subordinatorCoefficient}),
\begin{multline}
\notag
c(n) = \int_0^{1} \frac{1-\e^{-n y}}{n}\, {m} (\d y) + \int_1^{\infty} \frac{1-\e^{-n y}}{n}\, {m} (\d y) \\
\ge \int_0^{1} \frac{1-\e^{-n y}}{n}\, y\, {m} (\d y) + \int_1^{\infty} \frac{1-\e^{-n y}}{n}\, {m} (\d y) \\ \sim \frac{1}{n} \int_0^\infty \min \{ 1, y \}\,{m} (\d y) \;\; \text{as $n \rightarrow \infty$}\,.
\end{multline}
Hence the continued fraction converges 
for every $E \in {\mathbb C} \backslash {\mathbb R}_+$, and  the limit $K_\infty$ is a function of $E$
analytic in ${\mathbb C} \backslash {\mathbb R}_+$.

Next, let us show that $E^n\hat{f}(n)$ is a recessive solution of the difference equation (\ref{ascendingDifferenceEquation}) when $E<0$:
since it is a solution, we may write
$$
E^n \hat{f} (n) = \alpha \left [ q_n K_\infty (E) -p_n \right ] + \beta q_n 
$$
for some constants $\alpha$ and $\beta$. Note in particular that the $q_n$ are positive, because the $c(n)$ are positive.
Suppose that $\beta \ne 0$. Then, since $q_n K_\infty - p_n$ is recessive,
we must have
$$
E^n \hat{f} (n) \sim \beta q_n \;\;\text{as $n \rightarrow \infty$}\,.
$$
But this is absurd because the right-hand side is of the same sign for every $n$ whereas,
since $E<0$ and $\hat{f}(n)$ is by construction positive, the left-hand side alternates
in sign. Hence $\beta = 0$ and the recessiveness of $E^n \hat{f}(n)$ follows from Remark~\ref{recessiveRemark}.

We deduce from Pincherle's Theorem that
$$
\Omega(E) = \frac{c(0)}{2} + K_{\infty}(E)
$$
for $E <0$, and thence for 
every $E \in {\mathbb C} \backslash {\mathbb R}_+$ by analytic continuation.

\begin{remark}
This representation of the complex Lyapunov exponent in terms of the continued fraction associated
with the difference equation for the Mellin transform is not valid for every L\'{e}vy 
process. For the Brownian motion with drift of Example \ref{brownianWithDriftExample}, $\hat{f}(s)$
is {\em not} recessive, and the continued fraction converges to a ratio of $I_{\nu}$.
\label{continuedFractionRemark}
\end{remark}

\begin{remark}
Continued fractions have been used before in the study of other one-dimensional disordered systems--- most notably by Nieuwenhuizen \cite{Ni}
and Nieuwenhuizen \& Luck \cite{NL}. In these works, however, the continued fraction arose only in very special cases where the noise or
randomness could be modelled in terms of the exponential and other closely related distributions. 
\label{otherContinuedFractionRemark}
\end{remark}

\subsection{The density of states in the subordinator case}
\label{stieltjesSubsection}
In particular,
we have the formula
\begin{equation}
  N(E)  =  \frac{1}{\pi} \,\text{Im}\left[\frac{u_1}{u_0}\right]
  = - \frac{1}{\pi} \,\text{Im} \left[ K_{\infty}(E+\i 0+) \right] 
  \,, \;\; E>0\,.
  \label{densityOfStatesFormula}
\end{equation}
On the other hand, since the $c(n)$ are all positive, the limit of the continued fraction is the Stieltjes transform
of a certain measure, say $\kappa$, on the half-line ${\mathbb R}_+$ and we may write (see for instance \cite{BO,NS})
\begin{equation}
-\frac{K_{\infty}(E)}{E} = \int_0^\infty \frac{\kappa (\d t)}{t-E}\,, \;\; E \in {\mathbb C} \backslash {\mathbb R}_+\,.
\label{stieltjesTransform}
\end{equation}
There is a simple relationship between the Stieltjes measure $\kappa$
and the 
integrated density of states $N$: by the Stieltjes--Perron inversion
formula, for every open interval $(a,b) \subset {\mathbb R}_+$
$$
\kappa ( a,b ) = -\frac{1}{\pi} \int_a^b \text{Im}  \left [ \frac{K_{\infty} (t + \i 0+)}{t} \right ] \,\d t\,.
$$
We then deduce from Equation (\ref{densityOfStatesFormula})
\begin{equation}
\kappa (\d t) = N(t) \frac{\d t}{t}\,.
\label{stieltjesDensityOfStates}
\end{equation}
Now, as is well-known, the coefficients $c(n)$ appearing in the continued fraction $K$ may be expressed in terms of the positive integer moments of the
measure $\kappa$; see \cite{St} for detailed formulae.
So the problem of finding the density of states is essentially equivalent to a Stieltjes moment problem. 

\section{A new solvable case}
\label{hermiteSection}
Having reviewed in \S \ref{reviewSection} the few examples where the complex Lyapunov exponent is known explicitly, we now look for new cases.

\subsection{Masson's difference equation}
\label{massonSubsection}
Masson's catalog
of the solutions of the difference equation
\begin{equation}
v_{n+1} - (z-D n) v_n - ( A n^2 + B n + C ) v_{n-1} = 0\,, \quad n \in {\mathbb N}\,,
\label{massonDifferenceEquation}
\end{equation} 
expressed in terms of the hypergeometric function or its confluent limits, is particularly valuable in this respect \cite{Ma}. 
The correspondence between Masson's difference equation and
(\ref{ascendingDifferenceEquation}) is as follows:
set 
$$
c_n := c(n)
$$
and
\begin{equation}
u_n = \left ( \sqrt{-E}^n \prod_{j=0}^{n-1} c_j \right ) v_n\,.
\label{uAndv}
\end{equation}
Then $u_n$ solves the difference equation (\ref{ascendingDifferenceEquation}) if and only if $v_n$ solves the difference equation
$$
v_{n+1} - \frac{1}{\sqrt{-E}} v_n - \frac{1}{c_{n-1} c_n} v_{n-1} = 0\,.
$$
This is Masson's difference equation (\ref{massonDifferenceEquation}) with $D=0$ and $z=1/\sqrt{-E}$ provided that
\begin{equation}
c_{n-1} c_n = \frac{1}{A n^2 + B n + C}\,.
\label{massonCorrespondence}
\end{equation}
There are two possibilities: either
$$
c(n) \sim \frac{1}{\sqrt{A} \,n} \quad \text{or} \quad c(n) \sim \frac{1}{\sqrt{B n}} \quad \text{as $n \rightarrow \infty$}\,.
$$
The first case leads to solutions expressible in terms of the hypergeometric function; this is our earlier Example \ref{exponentiallyDistributedJumpsExample}.
The second case, developed in the next subsection, leads to solutions expressible in terms of parabolic cylinder functions; we follow
Masson's terminology and call it the ``Hermite case''.

\subsection{The Hermite case}
\label{hermiteSubsection}
Let the L\'{e}vy process be the subordinator with L\'{e}vy measure
\begin{equation}
  m(\d y) = \frac{2 p }{\sqrt{\pi}} \,
  \frac{\d}{\d y} \left [ \frac{-\e^{-q y}}{\sqrt{1-\e^{-2y}}} \right ]\,\d y
  \,, \quad q > 0\,.
  \label{parabolicLevyMeasure}
\end{equation}
The only dimensional parameter in this expression is $p$, which has the dimension of an inverse length, or equivalently of $\sqrt{E}$.
For convenience, we shall for the present set it to unity, and reintroduce it later whenever it becomes relevant.
To work out the corresponding L\'{e}vy exponent, introduce the {\em tail} of the L\'{e}vy measure:
$$
m(y,\infty) := 
\int_y^\infty {m}(\d t) = \frac{2}{\sqrt{\pi}} \frac{\e^{-q y}}{\sqrt{1-\e^{-2y}}}\,.
$$
This makes it clear that ${m}$ is not a finite measure and that 
\begin{equation}
m(\d y)\sim \frac{\d y}{y^{3/2}}  \quad \text{as } y \rightarrow 0+\,. 
\label{behaviourAtZero}
\end{equation}
We have
\begin{multline}
\notag
\Lambda(\theta) = \int_0^\infty \left( \e^{\i \theta y}-1 \right) \,m(\d y) 
= 
 \i \theta \int_0^\infty \d y \,\e^{\i \theta y} \, m(y,\infty) \\
= \frac{2\i\theta}{\sqrt{\pi}} \int_{0}^\infty \d y \,\e^{\i\theta y} \frac{\e^{-q y}}{\sqrt{1-\e^{-2y}}}
 \overset{\underset{\downarrow}{t=\e^{-y}}}{=}
  \frac{2\i\theta}{\sqrt{\pi}} 
  \int_{0}^1 \d t \,\frac{t^{q-\i\theta-1}}{\sqrt{1-t^2}} 
  = \frac{\i\theta}{\sqrt{\pi}} 
  {\tt B} \left (\frac{q-\i\theta}{2},\frac{1}{2} \right )\,.
\end{multline}
Here, we have made use of Formula 1 in \S 8.380 of \cite{GR}. 
Hence
$$
  \Lambda(\theta) = \i\theta 
  \frac{\Gamma\left( \frac{q-\i \theta}{2}\right)}
  {\Gamma\left( \frac{q-\i \theta+1}{2}\right)}
  \,.
$$
We note for future reference the large $\theta$ behaviour
\begin{equation}
  \label{eq:LevyExponentHermiteLarge}
  \Lambda(\theta) \sim -\sqrt{-2\i\theta}  \;\; \text{as $\theta \rightarrow \infty$}
\end{equation}
which is directly related to the small $y$ behaviour of the L\'{e}vy measure in Equation (\ref{behaviourAtZero}).
The coefficients appearing in the difference equation (\ref{ascendingDifferenceEquation}) are given by
\begin{equation}
  c(n) = c_n :=
  \frac{\Gamma\left( \frac{n+q}{2}\right)}{\Gamma\left( \frac{n+q+1}{2}\right)}
  \,.
\label{parabolicCoefficient}
\end{equation}
Hence
$$
\frac{1}{c_{n-1} c_n} = \frac{n+q-1}{2}\,.
$$

\begin{figure}[!ht]
  \centering
  \includegraphics[scale=0.65]{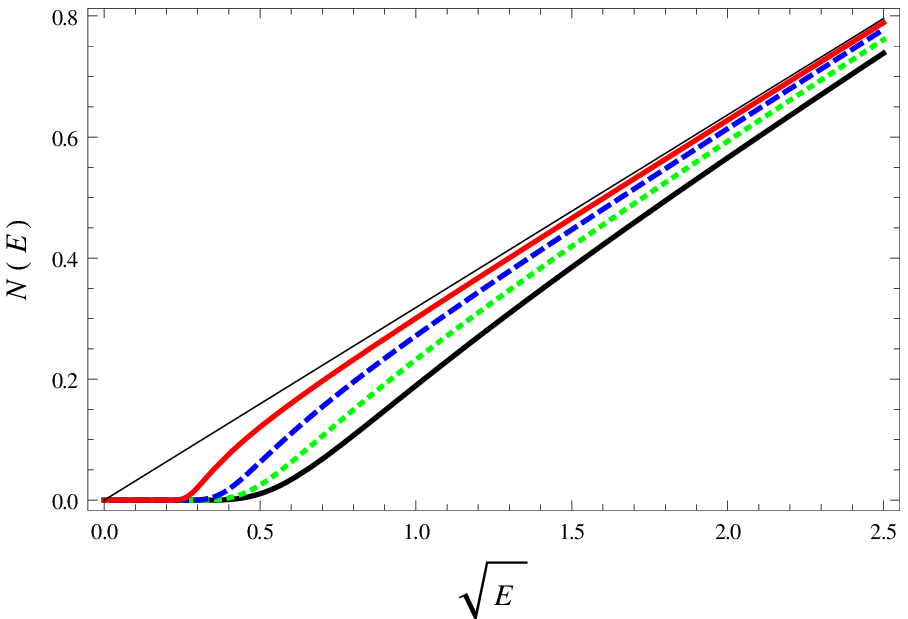}
  \hspace{0.25cm}
  \includegraphics[scale=0.65]{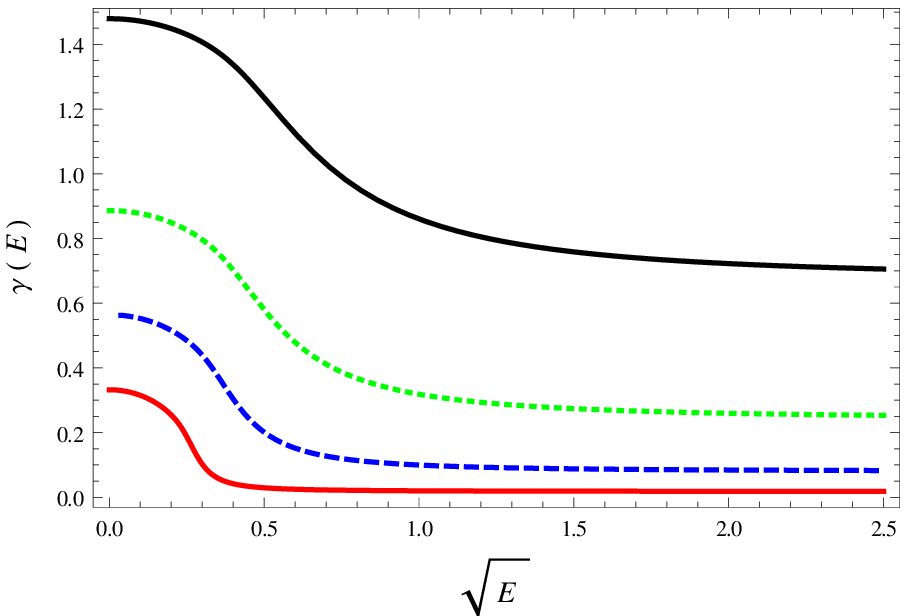}
  \caption{\it Integrated density of states and Lyapunov exponent for the Hermite case with $q=0.2$ (black continuous line), $0.5$ (green dotted line), $1$ (blue dashed), $2$ (red dashed-dotted) and $5$ (black continuous).}
  \label{densityOfStatesFigure}
\end{figure}

Let $u_n$ and $v_n$ be related
by Equation (\ref{uAndv}). Then Equation (\ref{massonCorrespondence}) says that $u_n$ solves the difference equation 
(\ref{ascendingDifferenceEquation})
if and only if $v_n$ solves
Masson's difference equation with
$$
z = \frac{1}{\sqrt{-E}},\;\;A = 0,\;\; B = \frac{1}{2},\;\; C = \frac{q-1}{2} \;\;\text{and} \;\; D = 0\,.
$$
Masson's Theorem 4 \cite{Ma} then says that the difference equation (\ref{ascendingDifferenceEquation}) has two linearly independent solutions $u_n^{\pm}$ given by
\begin{equation}
u_n^{\pm} := \left [ \mp \sqrt{-E/2} \right ]^n \frac{\Gamma \left ( n + q \right )}{\Gamma \left ( \frac{n + q}{2} \right )}
\, D_{-n-q} \left ( \pm \sqrt{-2 /E} \right )\,.
\label{hermiteSolution}
\end{equation}
This fact is easily verified by using the recurrence relation satisfied by the parabolic cylinder functions (see \cite{Er}, Vol. 2, Chapter VIII):
$$
D_{\nu+1} (x) - x D_{\nu}(x) + \nu D_{\nu-1} (x) = 0\,.
$$
Furthermore, $u_n^+$ is recessive. By using the identity
$$
\Gamma ( 2 z) = \frac{1}{\sqrt{\pi}} 2^{2 z-1} \Gamma (z) \,\Gamma \left ( z + \frac{1}{2} \right )
$$
we deduce the following expression for the Mellin transform of the invariant density:
\begin{equation}
\hat{f} (s) = \left [ \sqrt{-2 / E} \right ]^{s} 
\frac{\Gamma \left ( \frac{s + q+1}{2} \right )}{\Gamma \left ( \frac{q+1}{2} \right )}
\, \frac{D_{-s-q} \left ( \sqrt{-2/ E} \right )}{D_{-q} \left ( \sqrt{-2/ E} \right )}\,.
\label{hermiteMellinTransform}
\end{equation}
Therefore, after re-introducing the parameter $p$, we
obtain
\begin{equation}
  \Omega(E) = 
  \frac{p}{2} 
  \frac{\Gamma\left( \frac{q}{2} \right)}{\Gamma\left( \frac{q+1}{2} \right)} 
  + \frac{q\sqrt{-E}}{\sqrt{2}} \frac{\Gamma \left ( \frac{q}{2} \right )}{\Gamma \left ( \frac{q+1}{2} \right )}
\,\frac{D_{-q-1} \left ( p \sqrt{-2 / E} \right )}{D_{-q} \left ( p \sqrt{-2 /E} \right )}\,.
\label{hermiteCharacteristicFunction}
\end{equation}
Using $D_{-q}(0)=2^{-q/2}\sqrt\pi/\Gamma(\frac{1+q}{2})$ we obtain
\begin{equation}
\Omega(E) \sim \gamma_\infty + \sqrt{-E} 
\quad \text{as $E \rightarrow -\infty$}
  \:.
\end{equation}
The constant term is given by 
\begin{equation}
  \gamma_\infty = p\frac12\, c(0)
  + p \frac{2^q}{\sqrt\pi} 
  \frac{\Gamma(\frac{q+1}{2})^2-\frac q2\Gamma(\frac{q}{2})^2}{\Gamma(q)}
\end{equation}
where
$$
c(0)={\mathbb E}\big( W(1))
=p\frac{\Gamma\left( \frac{q}{2} \right)}{\Gamma\left( \frac{q+1}{2} \right)}
  \:.
$$
An analytic continuation to positive energies shows that
$$
 \gamma_\infty = \lim_{E\to+\infty}\gamma(E)
  \:.
$$
%
%
%
The Stieltjes measure of a ratio of parabolic cylinder functions is given explicitly in Ismail \& Kelker \cite{IK}, Theorem 1.4. That result,
combined with Equation (\ref{stieltjesDensityOfStates}), leads to the
following explicit formula for the integrated density of states:
\begin{equation}
N(E) = \frac{1}{2 \sqrt{\pi}} \frac{\Gamma \left ( \frac{q}{2} \right )}{\Gamma (q) \Gamma \left ( \frac{q+1}{2} \right )}
\frac{\sqrt{E}}{\left | D_{-q} \left ( \i p \sqrt{2/ E} \right ) \right |^2}\,.
\label{hermiteDensityOfStates}
\end{equation}
The low energy behaviour is 
\begin{equation}
N(E) \sim \frac{p^{2 q}}{\Gamma \left ( \frac{q+1}{2} \right )^2}
\,E^{\frac{1}{2}-q} \,\e^{-p^2/E} \quad \text{as $E \rightarrow 0+$}\,.
\label{hermiteLowEnergy}
\end{equation}
Plots of $N$ and $\gamma$ for various values of the parameter $q$ are shown in Figure \ref{densityOfStatesFigure}.

\begin{remark}
Let $E<0$ and set $k = \sqrt{-E}$. By using Equation (7.727) in \cite{GR}, we can invert the Mellin transform $\hat{f}$ to obtain
\begin{equation}
f(z) = C(k,q) \frac{z^{1-q}}{\left ( z^2-k^2 \right )^{\frac{3}{2}}} \,\exp \left ( - \frac{z^2/k^2}{z^2-k^2} \right ) {\mathbf 1}_{(k,\infty)}(z)
\label{hermiteDensity}
\end{equation}
where
$$
C(k,q) := 2 \left ( \frac{k}{\sqrt{2}} \right )^q \frac{\e^{\frac{1}{2k^2}}}{\Gamma \left ( \frac{q+1}{2} \right ) D_{-q} \left ( \frac{\sqrt{2}}{k} \right )}\,.
$$
\label{hermiteRemark}
\end{remark}

\begin{remark}
The limit $q \rightarrow 0+$ is singular, since
$$
{\mathbb E} \left ( W(1) \right ) =  c(0) = p \frac{\Gamma \left ( \frac{q}{2}\right )}{\Gamma \left ( \frac{q+1}{2}\right )} \rightarrow \infty
\quad \text{as $q \rightarrow 0+$}\,.
$$
On the other hand,
if we set $q=0$ in Equation (\ref{parabolicLevyMeasure}), we obtain the L\'{e}vy measure of the process with exponent
$$
\Lambda ( \theta) = \frac{2 p}{\sqrt{\pi}} + \i p \,\theta \frac{\Gamma \left ( \frac{- \i \theta}{2} \right )}{\Gamma \left ( \frac{1- \i \theta}{2} \right )}\,.
$$
This process is well-behaved; indeed
$$
c(n) = p \left [ \frac{\Gamma \left ( \frac{n}{2}\right )}{\Gamma \left ( \frac{n+1}{2}\right )} - \frac{2}{\sqrt{\pi} n} \right ]
$$
and so 
$$
{\mathbb E} \left ( W(1) \right ) = c(0) = \frac{p}{\sqrt{\pi}} 2 \ln 2 \quad \text{for $q=0$}\,.
$$
Unfortunately,
Equation (\ref{massonCorrespondence}) does not hold in this case, and so we cannot use Masson's results.
\label{qIsZeroRemark}
\end{remark}

\begin{remark}
\label{ornsteinUhlenbeckRemark}
One reason for the importance of the subordinator subclass of the L\'{e}vy processes is that they arise
naturally in the study of diffusion processes. More precisely, let $X = X(t)$ be a diffusion process on $\mathbb{R}$, started at the origin,
and driven by the Langevin equation
$$
\dot{X} = {\mathtt v} \left ( X \right ) + \eta
$$ 
where $\eta$ is white noise and $\mathtt v$ is the deterministic drift function. Then the {\em inverse local time} of $X$ is a process, say $\tau_t$, that measures how much time elapses before $X$ has spent
a total time $t$ in the vicinity of its starting point. It is a well-known fact that
$\tau$ is a subordinator. Given the drift $\mathtt v$, one may in principle compute the L\'{e}vy exponent of the inverse local time by studying a certain
Sturm--Liouville problem associated with the diffusion. In particular, for the
drift
$$
{\mathtt v} (x) = -x\,,
$$
which corresponds to the familiar Ornstein--Uhlenbeck diffusion process, one finds 
that the L\'{e}vy measure of $\tau$
is exactly the measure $m$ in Equation (\ref{parabolicLevyMeasure}) for $q=1$; see for instance \cite{HT}.
It is unclear whether this intriguing coincidence is the manifestation of a deeper connection between the Ornstein--Uhlenbeck
process and the disordered model.
\end{remark}

\section{The low-energy behaviour of the density of states}
\label{semiclassicalSection}
Throughout this section, we suppose that the L\'{e}vy process $W$ is a subordinator.
Whereas Kotani's study rested on a semiclassical analysis of a differential equation, we must,
in the supersymmetric case,
work with a difference equation.
Finite-difference versions of the WKB method were expounded by Dingle \& Morgan \cite{DM}
and Braun \cite{Br}, and we shall draw on their work in what follows.
It will be convenient to recast the difference equation (\ref{ascendingDifferenceEquation}) in the
form  
\begin{equation}
y_{n+1} + y_{n-1} = \frac{c(n)}{\sqrt{E}} \,y_n 
\label{dingleDifferenceEquation}
\end{equation}
obtained by setting
$$
u_n = E^{\frac{n}{2}} y_n\,.
$$
The basic formula (\ref{densityOfStatesFormula}) for the
integrated density of states becomes 
\begin{equation}
N(E) = \frac{\sqrt{E}}{\pi} \,\text{Im} \left [ \frac{y_1(E+\i 0+)}{y_0 (E+ \i 0+)} \right ]
\label{dingleDensityOfStatesFormula}
\end{equation}
where $y_n(E)$ is any recessive solution of Equation (\ref{dingleDifferenceEquation}).

\subsection{The remainders}
\label{remaindersSubsection}
The starting point of our analysis is the represention of the recessive solution in terms of the remainders
\begin{equation}
z_n := \cfrac{E}{c(n) + \cfrac{-E}{c(n+1) + \cfrac{-E}{c(n+2) + \cdots}}}\,.
\label{remainders}
\end{equation}
These remainders are certainly well-defined numbers for $E \notin {\mathbb R}_+$ because the continued fraction on the right-hand side converges.
We note the obvious identity
\begin{equation}
z_n = \frac{E}{c(n)-z_{n+1}}\,.
\label{remainderIdentity}
\end{equation}
Set
\begin{equation}
y_n^+ := \prod_{j=1}^n \frac{z_j}{\sqrt{E}}
\label{remainderSolution}
\end{equation}
If we take the empty product to equal $1$, it is easy to show by induction on $n$ that
$$
E^{\frac{n}{2}} y_n^+ = E^n \hat{f}(n) \;\; \text{for every $n \ge 0$}\,.
$$
This shows that $y_n^+$ is a recessive solution of the difference equation (\ref{dingleDifferenceEquation}).
We remark also that, since $c(n)$ decays to zero, 
$z_n$ and $z_{n+1}$ have a common limit, namely $-\sqrt{-E}$, as $n \rightarrow \infty$. Hence, replacing $z_{n+1}$ by
$z_n$ in Equation (\ref{remainderIdentity}) yields
\begin{equation}
z_n \sim \frac{c(n)-\sqrt{c(n)^2-4E}}{2} \quad \text{as $n \rightarrow \infty$}\,.
\label{remainderEstimate}
\end{equation}
We select this particular solution of the quadratic equation because it yields the correct behaviour of $z_n$
in the limit $E \rightarrow 0$; see Equation (\ref{remainderIdentity}).

Henceforth, we suppose that $E>0$ and set $k = \sqrt{E}$. The expression $y_n^+(E)$ should then be
interpreted as $y_n^+(E+\i0+)$.

\subsection{An alternative formula for the density of states}
\label{alternativeSubsection}
The difference equation (\ref{dingleDifferenceEquation}) has a
{\em turning point} $n_k$, defined to be 
the largest integer not exceeding the root of
$$
c(n) = 2 k \,.
$$
We proceed to express the integrated density of states in terms of the recessive solution at the
turning point.

Let $0 < E \le c(1)$. For $n > n_k$, we have
$$
\frac{c(n)+\i \sqrt{4 k^2-c(n)^2}}{2} = k \,\e^{\i \theta_n}
$$
where
$$
\theta_n = \arccos \frac{c(n)}{2k} > 0\,.
$$
It then follows from Equation (\ref{remainderEstimate}) that, for $n > n_k$ and $E>0$, there holds
\begin{equation}
y_n^+ \sim y_{n_k }^+ \exp \left ( \i \sum_{j>n_k} \theta_j \right ) \quad \text{as $n \rightarrow \infty$}\,.
\label{asymptoticSolution}
\end{equation}
Since $c(n)$ is real, the complex conjugate 
$$
y_n^- := \overline{y_n^+} 
$$ 
is also a solution of the difference equation (\ref{dingleDifferenceEquation}). Furthermore, $y_n^+$ and $y_n^-$ are linearly independent. 
Introduce their Wronskian
$$
{\mathcal W}_n := \begin{vmatrix}
y_{n-1}^- & y_n^- \\
y_{n-1}^+ & y_n^+
\end{vmatrix}\,.
$$
Then
$$
N(E) = \frac{k}{2 \pi} \,\text{Im} \left [ \frac{y_1^+}{y_0^+} - \frac{y_1^-}{y_0^-} \right ] = 
\frac{k}{2 \pi}  \,\text{Im} {\mathcal W}_1\,.
$$
On the one hand, 
by making use of the difference equation (\ref{dingleDifferenceEquation}), it is easy to see that the 
Wronskian of {\em any} two linearly independent solutions satisfies
$$
{\mathcal W}_{n} = {\mathcal W}_1
$$
for every $n \in {\mathbb N}$.
On the other hand, by construction, $y_n^+$ has the asymptotic behaviour (\ref{asymptoticSolution}).
Hence
\begin{equation}
\notag
{\mathcal W}_n \sim \left | y_{n_k}^+ \right |^2  \left [ \e^{\i \theta_n} - \e^{-\i \theta_n} \right ]
= \left | y_{n_k}^+ \right |^2 \left [ 2 \i \sin \theta_n  \right ]
\sim  \left | y_{n_k}^+ \right |^2 2 \i \quad \text{as $n \rightarrow \infty$}\,.
\end{equation}
We deduce
$$
{\mathcal W}_1 = \left | y_{n_k}^+ \right |^2 2 \i
$$
and so
\begin{equation}
N(E) = \frac{k}{\pi} \left | y_{n_k}^+ \right |^2\,.
\label{newDensityOfStatesFormula}
\end{equation}

\subsection{Semiclassical approximation}
\label{WKBsubsection}

There remains to find the small $k$ behaviour of $y_{n_k}^+$. From this point on,
we shall not aim at a completely rigorous treatment, but rather at
a simple informative estimate based on heuristic considerations.

Following Braun \cite{Br}, we introduce a small parameter $h>0$ such that
$$
h \rightarrow 0 \;\;\text{if and only if} \;\; k \rightarrow 0\,.
$$
The relationship between $h$ and $E$ is assumed to be such that there exist functions $C$ and $Z$
defined on ${\mathbb R}_+$
satisfying
\begin{equation}
C(nh) = \frac{c(n)}{2 k} \;\;\text{and}\;\; Z(nh) = \frac{z_n}{k}\,.
\label{Cfunction}
\end{equation}
Detailed examples will be given in due course.
Equation (\ref{remainderIdentity}) for the remainder may then be expressed as
\begin{equation}
2 C(x) - Z(x+h) = \frac{1}{Z(x)}
\label{remainderEquation}
\end{equation}
where 
$$
x = n h\,.
$$
We look for a solution of Equation (\ref{remainderEquation}) of the form
\begin{equation}
Z(x) = \exp \left [ S_0(x) + O \left ( h \right ) \right ] \quad \text{as $h \rightarrow 0$}\,.
\label{WKBexpansion}
\end{equation}
When we substitute into Equation (\ref{remainderEquation}), use the MacLaurin series of the exponential function, and
equate like powers of $h$, we find, for the leading term,
\begin{equation}
\frac{\e^{S_0}+\e^{-S_0}}{2} = C\,.
\label{controllingFactorEquation}
\end{equation}
Then, by making use of the Euler-MacLaurin formula, we obtain
\begin{equation}
y_n^+ = \prod_{j=1}^n Z(jh) = \exp \left \{ \frac{1}{h} \int_h^x S_0 (t)\,\d t + \frac{S_0(x)+S_0(h)}{2} + O(1)\right \}\,.
\label{WKBsolution}
\end{equation}
In order to have
$$
Z(n h) \sim \frac{E}{c(n)} \quad \text{as $k \rightarrow 0$}
$$
for $n < n_k$,
we must select the following solution of Equation (\ref{controllingFactorEquation}):
$$
S_0 (x) = -\text{arccosh} \left [ C(x) \right ]\,.
$$
Then
\begin{equation}
\notag
y_{n_k}^+  = \exp \left \{ -\frac{1}{h} \int_h^{x_k} \text{arccosh} \left [ C(t) \right ]\,\d t - \frac{1}{2} \text{arccosh} \left [ C(h) \right ] +  O(1) \right \}
\end{equation}
where $x_k$ is defined by
\begin{equation}
C(x_k)=1\,.
\label{turningPoint}
\end{equation}
In view of Formula (\ref{newDensityOfStatesFormula}), this leads to the estimate
\begin{equation}
\ln N = -\frac{2}{h} \int_h^{x_k} \text{arccosh} \left [ C(t) \right ]\,\d t - \text{arccosh} \left [ C(h) \right ] + \ln k + O(1)
\label{kotaniFormula}
\end{equation}
as $k \rightarrow 0$. The validity of this estimate is open to question, for we have assumed implicitly that the 
$O(1)$ correction term in Equation (\ref{WKBsolution}) is bounded uniformly for $h < x \le x_k$. We proceed to
discuss some examples where this assumption seems justified.

\begin{example}
For the L\'{e}vy process of Example \ref{exponentiallyDistributedJumpsExample}, 
set $h = k$.
Then
$$
\frac{c(n)}{2 k} = \frac{\rho}{2 \left ( nh + qh \right )}  
 =: C( nh)\,. 
$$
The undesirable dependence of the function $C$ on $h$ could be removed
by shifting the index $n$. As it is, however,
Formula (\ref{kotaniFormula}) makes sense and may be used directly; shifting the index
would eventually lead to the same result.
We have
\begin{multline}
\notag
- \frac{2}{h} \int_h^{x_k} \text{arccosh} \left [ \frac{\rho}{2 \left ( t + qh \right )} \right ] \,\d t \\
\overset{\underset{\downarrow}{y =C(t)}}{=} -\frac{\rho}{h} \left [ \frac{\text{arccosh}\, y}{y} + \arctan \frac{1}{\sqrt{y^2-1}} \right ] \Bigl |_{\frac{\rho}{2h(q+1)}}^1 \\
= -\frac{\rho}{h} \frac{\pi}{2} - 2 (q+1)\,\ln h + O(1)\,.
\end{multline}
We deduce
$$
N(E) \sim  A \,E^{-q} \,\exp \left [ -\frac{\pi}{2}\frac{\rho}{\sqrt{E}} \right ] \quad \text{as $E \rightarrow 0+$}
$$
for some constant $A$ independent of $E$.
This agrees with Equation (\ref{bienaime}).
\label{gammaExample}
\end{example}

\begin{remark}
We remark that, for every finite L\'{e}vy measure,
$$
c(n) := \int_0^\infty \frac{1-\e^{-ny}}{n} \,m (\d y) \sim \frac{1}{n} \int_0^\infty m( \d y) \quad \text{as $n \rightarrow \infty$}\,.
$$
Therefore the asymptotic relation
\begin{equation}
\ln N(E) \sim -\frac{\pi}{2}\frac{1}{\sqrt{E}} \int_0^\infty m(\d y) \quad \text{as $E \rightarrow 0+$}
\label{finiteBehaviour}
\end{equation}
holds more generally for every subordinator with a finite L\'{e}vy measure.
\label{finiteRemark}
\end{remark}

\begin{example}
For the solvable model of \S \ref{hermiteSubsection}, we have
$$
c(n) = \frac{\Gamma \left ( \frac{n+q+1}{2} \right )}{\Gamma \left ( \frac{n+q}{2} \right )}\,.
$$
Strictly speaking, the corresponding function $C$ is defined implicitly by Equation (\ref{Cfunction}), but a more tractable expression
may be obtained by using the identity
$$
c(n-1) \,c(n)  = \frac{2}{n+q-1}\,.
$$
Then
$$
C(n h)^2 = \frac{1}{2 k^2 \left ( n + q - \frac{1}{2} \right )} + O(h^2) \quad \text{as $h \rightarrow 0$}\,.
$$
So, for the purpose of our calculation, we may take
$$
h := 2 k^2 \;\;\text{and}\;\; C(x) := \frac{1}{\sqrt{x+h \left ( q - \frac{1}{2} \right )}}\,.
$$
Then
\begin{multline}
\notag
-\frac{2}{h} \int_h^{x_k} \text{arccosh} \left [ C(t) \right ]\,\d t \overset{\underset{\downarrow}{y =C(t)}}{=} - \frac{4}{h} \left \{ \frac{\sqrt{y^2-1}}{2 y} - \frac{1}{2 y^2} \text{arccosh}\, y \right \} \Bigl |_{\left [ h ( q+ \frac{1}{2} )\right ]^{-\frac{1}{2}}}^1 \\
\sim -\frac{2}{h} - \left ( q + \frac{1}{2} \right )\,\ln h + O(1) \quad \text{as $h \rightarrow 0$}\,.
\end{multline}
Formula (\ref{kotaniFormula}) then yields
$$
N (E) \sim A \,E^{\frac{1}{2}-q} \,\e^{-\frac{1}{E}} \quad \text{as $E \rightarrow 0+$}\,.
$$
This is in agreement with the exact result found in \S \ref{hermiteSubsection}; see Equation (\ref{hermiteLowEnergy}).
\label{hermiteExample}
\end{example}

\begin{example}
The so-called {\em alpha-stable subordinator} has for its L\'{e}vy measure
$$
m(\d y) = p\, \frac{\alpha }{\Gamma (1-\alpha)} \frac{\d y}{y^{1+\alpha}}\,, \quad 0 < \alpha < 1\,.
$$
The corresponding L\'evy exponent is
\begin{equation}
  \Lambda(\theta) = - p\, (-\i\theta)^\alpha
  \:.
  \label{eq:LevyExponentAlphaStable}
\end{equation}
Some of the properties of alpha-stable subordinators are reviewed in the appendix.

For $p=1$, the continued fraction coefficients are 
$$
c(n) = n^{\alpha-1}
$$
and so we set
$$
h :=  (2 k)^{\frac{1}{1-\alpha}}\;\;\text{and}\;\; C(x) := x^{\alpha-1}\,.
$$
Then
\begin{equation}
\notag
-\frac{2}{h} \int_h^{x_k} \text{arccosh}^{-1} \left [ C(t) \right ]\,\d t 
= -\frac{2}{h} \mu_\alpha - 2(1-\alpha)\,\ln h + O(1) \quad \text{as $h \rightarrow 0$}
\end{equation}
where we have set
\begin{equation}
  \notag
  \mu_\alpha := (1-\alpha) 
  \,{_2}F_1\!\left(
    \frac{1}{2} , \frac{1}{2(1-\alpha)} ; 1 + \frac{1}{2(1-\alpha)} ;
    1 
  \right)
  = \frac{\pi}
         {\text{\tt B}\left( \frac{1}{2} + \frac{1}{2(1-\alpha)},
             \frac{1}{2} \right)}
  \,.
\end{equation}
We deduce from Formula (\ref{kotaniFormula})
\begin{equation}
  N(E) \sim A \,E^{\frac{1}{2}} \,
  \exp \left[
     -\frac{\mu_\alpha}{2^\frac{\alpha}{1-\alpha}} \,
     (p^2/E)^{\frac{1}{2(1-\alpha)}} 
  \right] 
  \quad \text{as } E \rightarrow 0+\,.
  \label{eq:IDoSSusyAlphaStable}
\end{equation}
\label{stableSubordinatorExample}
\end{example}

\subsection{The case of infinite L\'evy measure}
\label{infiniteSubsection}
Let us now discuss the case 
$$
\int_{\mathbb{R}_+} m(\d y)=\infty
$$
in greater generality.
A first interesting observation is that the two instances encountered earlier,
namely the Hermite case and the alpha-stable case with $\alpha=1/2$, have led to the same
leading behaviour of the integrated density of states at low energy; compare
Equations (\ref{hermiteLowEnergy}) and (\ref{eq:IDoSSusyAlphaStable}).
The mathematical explanation is that the corresponding L\'evy measures have the same singularity
at $y=0$; the fact that their tails are very different has little bearing on the low-energy
behaviour of $N$.
The physical interpretation is that, for an infinite L\'evy measure
$m$ with support $\mathbb{R}_+$, the low energy eigenstates of
the supersymmetric Hamiltonian
are strongly affected by the frequent small jumps but depend weakly on
the rare large jumps of the L\'evy process. 
This contrasts with the case of a finite measure, where
the exponent appearing in Formula (\ref{finiteBehaviour})
indicates that there is no such distinction.

Turning then to the analysis of the low-energy states of the
supersymmetric Hamiltonian,
consider the finite interval $[0,\ell]$. In the absence of any
potential, this interval supports an eigenstate of (kinetic) energy
$E_\mathrm{kin} \approx 1/\ell^2$.
A low-energy state $E \approx 1/\ell^2$ is possible if, in the interval $0 \le x \le \ell$,  the potential $V=w^2/4-w'/2$
remains sufficiently small to ensure that $V \ll E_\mathrm{kin}$. 
Because $W(x)$ is a subordinator, we expect that the potential's average value is roughly proportional to
the square of $W(\ell)/\ell$.
Hence,
if $W(\ell)$ does not exceed some small value, say $\varepsilon$, then the potential brings in a contribution
$V \approx (\varepsilon/\ell)^2$ to the energy.
Equating $\ell$ and $1/\sqrt{E}$, we deduce
\begin{equation}
  \ln N(E) \sim \ln {\mathbb P} \left ( W \left (1/\sqrt{E} \right )< \varepsilon \right )
\;\;\text{as $E \rightarrow 0+$}
  \:.\label{eq:HeuristicArgument}
\end{equation}

We argued earlier that, in this limit, only the frequent small jumps of the
process matter. With this in mind, let us begin with the case of a L\'{e}vy noise
whose measure exhibits the ``alpha-stable singularity''
$$
m(\d y) \sim  p\,\frac{\d y}{y^{1+\alpha}}\,, \; 0<\alpha<1\,, \quad \text{as $y \rightarrow 0+$}.
$$
In the appendix, we describe
the implications of this singularity for the form of the probability density function, say $f_{W} (\cdot\,;x )$, of the
random variable $W(x)$; we find
$$
f_W(u\,;x) = \frac{1}{(px)^{1/\alpha}} \, f_W \left ( \frac{u}{(px)^{1/\alpha}}\,; 1 \right )\,.
$$
Equation (\ref{lowDensityBehaviour}) in the appendix, which describes the small $u$ behaviour of this density, then leads to
\begin{equation}
  \label{eq:SpectralSingularityGammaProcessSusyInfiniteMeasure}
  \ln N(E) \sim 
  \ln \int_0^{\varepsilon} f_W \left ( u\,; 1/\sqrt{E} \right )\,\d u
  \sim 
  - E^{-\frac1{2(1-\alpha)}} \quad \text{as $E \rightarrow 0+$}\,.
\end{equation}
Modulo a constant factor on the right-hand side, these heuristic arguments extend our earlier result \eqref{eq:IDoSSusyAlphaStable}.

Finally, let us mention very briefly the limiting case
$\alpha \to 0$, where the L\'{e}vy measure satisfies
\begin{equation}
  m( \d y) \sim \frac{\d y}{y} \quad \text{as $y \rightarrow 0+$}\,.
  \label{eq:SingularMeasureMarginalCase}
\end{equation}
The corresponding process $W$ shares this singularity with the so-called gamma subordinator; see the appendix. By the same heuristic arguments as before,
we arrive once again at Equation (\ref{eq:HeuristicArgument}), and the result is
\begin{equation}
  \label{eq:SpectralSingularityGammaProcess}
  \ln N(E) \sim  - \frac{\ln(1/E)}{\sqrt{E}} 
  \quad \text{as $y \rightarrow 0+$}\,.
\end{equation}

\subsection{Comparison with the Kotani case $V=w$}
\label{comparisonSubsection}
It is interesting to compare these results with those obtained by Kotani \cite{Ko} for the random Schr\"{o}dinger equation
$$
-\psi'' + w\, \psi = E \psi\,.
$$
For the compound Poisson process of Example \ref{gammaExample}, Kotani
showed rigorously that
\begin{equation}
N(E) \sim A \exp \left [ - \frac{\rho \,\pi}{\sqrt{E}} \right ] \quad
\text{as }E \rightarrow 0
\,,
  \label{eq:Lifshits}
\end{equation}
where $\rho$ is defined as before by \eqref{finiteLevyMeasure}.
This is the well-known Lifshits singularity \cite{LGP}.
For the alpha-stable subordinator
of Example \ref{stableSubordinatorExample}, he found
\begin{equation}
N(E) \sim A \exp \left [ -
  \nu_{\alpha}
  \,E^{\frac{-1-\alpha}{2(1-\alpha)}} \right ] \quad \text{as } E \rightarrow 0
  \label{eq:IDoSSchrodAlphaStable}
\end{equation}
where
$$
  \nu_\alpha := \frac{\left [ \frac{p \alpha}{\Gamma (1-\alpha)} \right ]^\frac{1}{1-\alpha}}{\text{\tt B} \left ( \frac{1}{1-\alpha},\frac{1}{2}\right )}
  \,.
$$
Kotani also obtained exact expressions for the constant factor $A$; see
\cite{Ko}, Theorem 4.7. In the supersymmetric case, 
the constant factor may in principle be obtained by including the next
term in the WKB expansion of the remainder $Z$,  
but the calculations involved are tedious.

The asymptotic behaviour (\ref{eq:IDoSSchrodAlphaStable})
may be recovered by a heuristic argument similar to the one
  developed in the supersymmetric case.
  The main difference is that, whereas in the supersymmetric
  case $V = w^2/4-w'/2$ the L\'evy process is dimensionless, in the case $V=w$ it has
  the dimension of the reciprocal of length. Consider as before
  an interval $[0,\ell]$, and suppose that the L\'evy process does not increase beyond a certain threshold
  $\varepsilon$ in that interval. The contribution of the potential
  energy now reads $V \approx \varepsilon/\ell$.
  Therefore a state of low energy $E \approx E_\mathrm{kin} \approx 1/\ell^2$ is possible
  if $V\ll E_\mathrm{kin}$ i.e. $\varepsilon \ll1/\ell\approx\sqrt{E}$. Hence
$$
\ln N(E) \sim \ln {\mathbb P} \left ( W \left ( 1/\sqrt{E} \right )< \varepsilon \right )
\sim  - E^{-\frac{1+\alpha}{2(1-\alpha)}} 
\quad \text{as $E \rightarrow 0+$}\,.
$$
The different exponent, compared to Equation \eqref{eq:SpectralSingularityGammaProcessSusyInfiniteMeasure}, comes from the fact that the threshold $\varepsilon$ is now dimensional and scales with energy as $\sqrt E$.
The case
$m(\d y)\sim \d y/y$ for $y\to0+$ leads to the same behaviour as in the supersymmetric case; see Equation~(\ref{eq:SpectralSingularityGammaProcess}).

\subsection{Classical diffusion}
\label{diffusionSubsection}
The formal equivalence between supersymmetric quantum mechanics and classical diffusion is well-known and 
has proved fruitful in the study of diffusion in a random environment
\cite{BCGL,Le,LeDMonFis99,TexHag09}. The purpose of this subsection is to outline the most immediate
implications of our results for this field of study.

Consider a diffusing particle placed initially at a position $x_0$.
The Fokker-Planck equation for the probability density $P(\cdot\,;t)$ of its position at time $t$ is
$$
\derivp{}{t}P(x;t) = \derivp{}{x}\left [ \derivp{}{x} + {\mathtt v}(x) \right ] P(x;t)
-{\mathtt a}(x)P(x;t)
$$
where ${\mathtt v}$ and ${\mathtt a}$ are, respectively, the drift and the absorption (killing) rate
of the diffusion.
The transformation 
$$
P(x;t) = \e^{-\frac{1}{2} \int_0^x {\mathtt v}(y)\, \d y}\, \psi(x;t)
$$
leads to the Schr\"odinger equation
$$
-\derivp{}{t}\psi = H\,\psi
\quad \text{for } 
H := -\deriv{^2}{x^2} 
+ \frac{{\mathtt v}^2}{4} - \frac{{\mathtt v}'}{2} + {\mathtt a}
\:.
$$
When ${\mathtt v}$ and ${\mathtt a}$ are random, the expected value of $P(x_0;t)$ turns out to be independent
of the starting point $x_0$; it is related to the density of states $N'(E)$ associated with the disordered system via
$$
\overline{P (x_0;t)} = \int_0^\infty \d E\,N'(E)\,e^{-Et}
\:.
$$
By standard Tauberian arguments, the long-time behaviour of this expected density is therefore completely determined by the low-energy
behaviour of the density of states: 
$$
\ln N(E)\sim -E^{\frac{\nu}{\nu-1}} \; \text{as $E \rightarrow 0+$} \quad \Rightarrow \quad \ln \overline{P (x_0;t)} \sim -t^{\nu}
\;\text{as $t \rightarrow \infty$}\,.
$$

The Kotani case $V = w$ corresponds to the choice ${\mathtt v} = 0$ and ${\mathtt a}=w$.
Here, the decay of the return probability density
follows from the decay of the total probability due to
absorption.
For finite L\'{e}vy measures, the Lifshits singularity (\ref{eq:Lifshits})
leads to the well-known
exponent $\nu=1/3$ \cite{Lu}. On the other hand, 
when the L\'evy measure is infinite, the behaviour
\eqref{eq:IDoSSchrodAlphaStable} leads to the exponent 
$$
\nu = \frac{1+\alpha}{3-\alpha}
\:.
$$  

The supersymmetric case $V = w^2/4-w'/2$ corresponds to the choice ${\mathtt v} = w$ and ${\mathtt a} =0$. In this case, there is no absorption,
the total probability is conserved, and the decay of the return probability density is due to the expected drift of the L\'{e}vy process: 
$$
{\mathbb E} \left ( W(x) \right ) = x \int_0^{\infty} y \,m(\d y) = 2 a x\,.
$$
For a pure deterministic drift, $\nu=1$ and the return density decays exponentially. For a non-zero L\'{e}vy measure, however,
the fluctuations in the drift bring about a slower rate of decay. For example, in the case of a finite measure, we find, as in Kotani's case,
$\nu=1/3$; for the alpha-stable case, we find 
$$
\nu = \frac{1}{3-2 \alpha}\,.
$$

For the singular measure \eqref{eq:SingularMeasureMarginalCase}, the spectral singularity \eqref{eq:SpectralSingularityGammaProcess} leads to the probability decay 
$$
\ln \overline{P (x_0;t)} \sim - t^{1/3} \ln^{2/3}t
\:.
$$
In this marginal case, the singular nature of the measure is only reflected in the logarithmic correction.
This behaviour holds for both the random drift case and the random absorption case.

\section{Conclusion}
\label{conclusionSection}

In this paper, we have developed the supersymmetric version of the Frisch--Lloyd methodology for computing
the density of states of a system with L\'{e}vy disorder. The main novelty is that, in the supersymmetric case,
the complex Lyapunov exponent may be expressed in terms of the positive solution of a {\em difference} equation. 
When the L\'{e}vy process is non-decreasing, 
the complex Lyapunov exponent has a continued fraction expansion whose coefficients have simple expressions in terms
of the L\'{e}vy exponent, and the problem of computing the density of states reduces to a Stieltjes moment problem.
One pleasing outcome was the discovery of a new solvable case.
Our efforts to adapt to the supersymmetric case
Kotani's semiclassical analysis of the low-energy regime have also borne some fruit, although our approach only gives the leading term. 
This analysis has shown in particular that, for infinite L\'{e}vy measures, the low-energy behaviour of the integrated density of states 
is controlled by the small jumps of the process. This result is in some sense the dual of that obtained in an earlier study by one of us, namely
that it is the tail of the L\'{e}vy measure that determines the high-energy regime \cite{Bi,BT}.

The paper raises a number of questions that deserve further study:
\begin{enumerate}
\item Much of the work presented here has been concerned with the subordinator case. From the
point of view of localisation, however, the most interesting processes are those with zero means, such as Brownian motion.
Hence we require a less restrictive theory.

\item As mentioned earlier, exponentials of L\'{e}vy processes may be viewed as the zero-energy case of the supersymmetric model. There has been
a great deal of recent work on such exponentials, including their close connection with self-similar processes \cite{Pa}
and a generalisation to {\em complex} exponentials \cite{GDL}.
Our own findings
encourage us to seek interpretations of some of these results in terms of disordered systems.
\end{enumerate}

\appendix

\section{Some L\'evy processes}
\label{levyAppendix}
We recall in this appendix some basic facts concerning the distributional properties of two important classes
of L\'{e}vy processes.

\subsection{The alpha-stable process}

We study the subordinator characterised by the infinite L\'evy measure
$$
m(\d y) = \frac{\alpha }{\Gamma (1-\alpha)} \frac{\d y}{y^{1+\alpha}}\,, \quad 0 < \alpha < 1\,.
$$
The fact that the L\'evy exponent 
$$
\Lambda(\theta)=-(-\i\theta)^\alpha
$$ 
has a simple power law indicates that the law is
stable, namely,
$$
W(x)\eqlaw x^{1/\alpha}W(1)
$$ 
or, in other terms,
that the probability density function $f_W(\cdot\,;x)$ of the random variable $W(x)$ satisfies
$$
f_W(u\,;x) = x^{-1/\alpha}\,f_W(x^{-1/\alpha} u\,;1 )
$$ 
where
\begin{equation}
  f_W(u\,;1) = \int_{\mathbb{R}}\frac{\d z}{2\pi}\,\e^{\i zu-(\i z)^\alpha}\,.
  \label{eq:LevyLaw}
\end{equation}
This distribution is a particular case of a more general two-parameter family of
L\'evy stable laws with probability density function
$$
  \mathcal{L}_{\alpha,\beta}(u)=
  \int_{\mathbb{R}}\frac{\d z}{2\pi}\,
  e^{\i zu-|z|^\alpha(1+\i\beta\,\mathrm{sign}(z)\tan[\pi\alpha/2])}
  \:.
$$
The parameter $\beta\in[-1,+1]$ tunes the asymmetry of the law;
$\beta=0$ corresponds to a symmetric law, such as the Cauchy
$\mathcal{L}_{1,0}$
or the Gaussian $\mathcal{L}_{2,0}$ laws,
while $\beta=+1$
corresponds to subordinators, whose distributions have support in $\mathbb{R}_+$ \cite{{BouGeo90,Zo}}.
The precise relationship between $f_W(\cdot\,;1)$ and ${\mathcal L}_{\alpha,\beta}$ is
$$
f_{W}(u\,;1)=[\cos(\pi\alpha/2)]^{-1/\alpha}\,
\mathcal{L}_{\alpha,1} \left ( [\cos(\pi\alpha/2)]^{-1/\alpha}u \right )
\:.
$$

The distribution \eqref{eq:LevyLaw} may be analysed as follows:
for $u<0$, the analyticity of the integrand in the lower
complex plane implies $f(u)=0$.
For $u>0$ we can deform the contour of integration in order to follow the
branch cut along the positive imaginary semi-axis; the result is
\begin{equation}
  f_W(u\,;1) = \int_0^\infty\frac{\d t}{\pi} 
  \sin\left ( t^\alpha\sin\pi\alpha\right )\,
  \,\e^{-tu-t^\alpha\cos\pi\alpha}
  \label{eq:2}
\end{equation}
and so we deduce an asymptotic power law tail similar to the L\'evy measure
\begin{equation}
  f_W(u\,;1) \sim \frac{\alpha}{\Gamma(1-\alpha)} u^{-1-\alpha} \quad \text{as $u \rightarrow \infty$}\,.
  \label{eq:Limitf2}  
\end{equation}

The behaviour of the density function \eqref{eq:LevyLaw} 
for small $u$ may be studied by
using a stationary phase approximation.
Let us write
$$
  f_W \left ( \lambda^{1-1/\alpha}\,;1 \right ) =  \lambda^{1/\alpha}
  \int_{\mathbb{R}}\frac{\d z}{2\pi}\,\e^{\lambda\,\varphi(z)}
  \;\;\text{ where }\;\;  \varphi(z)=\i z-(\i z)^\alpha
  \:.
$$
Since $\mathrm{Re}[\varphi(z)]<0$ for $\mathrm{Im} \,z<0$, we can
deform the contour of integration in order to pass through the
stationary point $z_0=-\i\alpha^{1/(1-\alpha)}$ such that $\varphi'(z_0)=0$;
this leads to
\begin{equation}
f_W (u\,; 1) \sim
\frac1{\sqrt{2\pi  B_\alpha}}\,
  u^{-\frac{2-\alpha}{2(1-\alpha)}}\,
  \exp\left[-A_\alpha u^{-\frac{\alpha}{1-\alpha}}\right] \;\;\text{as $u \rightarrow 0+$}
\label{lowDensityBehaviour}
\end{equation}
where $A_\alpha=(1-\alpha)\alpha^{\alpha/(1-\alpha)}$
and $B_\alpha=(1-\alpha)\alpha^{-1/(1-\alpha)}$.

For example, we can verify that these behaviours are the correct ones in  the case $\alpha=1/2$, since 
$\mathcal{L}_{1/2,1}(u)$, the density of the so-called
L\'evy distribution, admits the simple expression  
$$
f_W(u\,;1)=\frac{1}{2\sqrt{\pi}}\,u^{-3/2} \e^{-1/(4u)} \mathbf{1}_{(0,\infty)}(u)
\:.
$$

\subsection{The gamma process}

The gamma process is characterised by the infinite L\'evy measure
\cite{Ap1}
$$
m( \d y ) = \frac{\d y}{y} e^{-by}
\:.
$$
Using 
$$
\int_0^\infty\frac{\d t}{t}(\e^{-At}-\e^{-Bt})=\ln \frac{B}{A}
$$
we obtain
the L\'evy exponent
\begin{equation}
  \Lambda(\theta)  =  - \ln \left( 1 - \i \theta / b \right)
  \:.
\end{equation}
From the point of view of the statistical properties of the frequent small jumps, the gamma process can be understood as the $\alpha\to0$ limit of the alpha-stable process.

The distribution can be studied by considering the Fourier transform 
$$
f_W(u\,;x) = \int_{\mathbb{R}}\frac{\d z}{2\pi}\,
\frac{ \e^{\i z u} }{ (1+\i z/b)^x }
\:.
$$
A contour deformation leads to
\begin{equation}
  \label{eq:DistributionGammaProcess}
  f_W(u\,;x) = 
  \frac{ b^x }{ \Gamma(x) }\, u^{x-1} \, \e^{-b u} \,\mathbf{1}_{(0,\infty)}(u)
  \:. 
\end{equation}
In particular, the moments are finite, given by
$$
\mathbb{E}\left( W(x)^n \right)
  = \frac{\Gamma(n+x)}{\Gamma(x)}\,b^{-n}
\:.
$$

\bibliographystyle{amsplain}

\end{document}